\documentclass[review]{elsarticle}

\usepackage{bm,hyperref,xcolor,subfig}
\usepackage{scrextend}

\journal{Applied Mathematics and Computation}

\begin{document}

\begin{frontmatter}

\title{Numerical evolution of the resistive relativistic magnetohydrodynamic equations: a minimally implicit Runge-Kutta scheme}


\author[mates]{Isabel Cordero-Carri\'on\corref{mycorrespondingauthor}}
\cortext[mycorrespondingauthor]{Corresponding author}
\ead{isabel.cordero@uv.es}

\author[mates]{Samuel Santos-P\'erez}
\author[mates]{Clara Mart\'inez-Vidallach}

\address[mates]{Departamento de Matem\'aticas, Universitat de Val\`encia, Calle Doctor Moliner 50, E-46100 Burjassot, Val\`encia, Spain}

\begin{abstract}
We present the Minimally-Implicit Runge-Kutta (MIRK) methods for the numerical evolution of the resistive relativistic magnetohydrodynamic (RRMHD) equations, following the approach proposed by Komissarov (2007) of an augmented system of evolution equations to numerically deal with constraints. Previous approaches rely on Implicit-Explicit (IMEX) Runge-Kutta schemes; in general, compared to explicit schemes, IMEX methods need to apply the recovery (which can be very expensive computationally) of the primitive variables from the conserved ones in numerous additional times. Moreover, the use of an iterative process for the recovery could have potential convergence problems, increased by the additional number of required loops. In addition, the computational cost of the previous IMEX approach in comparison with the standard explicit methods is much higher. The MIRK methods are able to deal with stiff terms producing stable numerical evolutions, minimize the number of recoveries needed in comparison with IMEX methods, their computational cost is similar to the standard explicit methods and can actually be easily implemented in numerical codes which previously used explicit schemes. Two standard numerical tests are shown in the manuscript.
\end{abstract}


\end{frontmatter}


\section{Introduction}
Significant magnetic fields are present in some relevent astrophysical scenarios, like accretion disks, active galactic nuclei, relativistic jets, quasars, or compact objects. See, for example, references \cite{BalbusHawley,BlaesBalbus,DeVilliersetal,Fragileetal,Gableretal,KaspiBeloborodov,KomissarovPorth,Marti,Perucho,Spruit} for some general reviews.

If we consider numerical simulations solving the ideal magnetohydrodynamic (MHD) equations \cite{Palenzuela2009}, effects induced by numerical errors and numerical resistivity will appear. These effects depend on the numerical method and resolutions used, and the physical resistivity is therefore not modeled consistently. A consistent treatment for the resistivity is necessary.

In the following, Greek indices ($\mu$, $\nu$, $\xi$, \ldots) are used for the 4 tensors and run from 0 to 3. Latin indices ($i$, $j$, $k$, \ldots) are used for the 3-spatial tensors and run from 1 to 3. We set the speed of light $c=1$. We also use Einstein's summation convention over repeated indices.

\section{Structure of the evolution system of equations}
We consider $n^{\mu}$ to be the time-like translational Killing vector field in a flat (Minkowski) space-time, so in this case $n_\mu=(-1,0,0,0)$ and the Levi-Civita symbol $\epsilon^{\mu\nu\xi}$ is non-zero only for spatial indices.

In the case of the resistive relativistic MHD (RRMHD) equations \cite{Komissarov2007}, we have to deal with a hyperbolic system of evolution equations for the rest mass-energy density of matter $\rho$, the components of the velocity field measured by the inertial observer $v^i$, the specific internal energy density $\epsilon$, the electric charge density $q$, the components of the electric field $E^i$ and the components of the magnetic field $B^i$. In addition, we have two constraint equations: the divergence of the magnetic field has to vanish and the divergence of the electric field equals the electric charge density. Shock and rarefaction waves potentially develop in the evolution of these equations, even when one starts from smooth initial data, and therefore high resolution shock capturing (HRSC) methods \cite{Anton2006} must be used in order to properly capture these phenomena. However, in this manuscript we will consider only smooth initial data and smooth data during the evolution, and we will focus on how to deal with the resistive source terms numerically, which become stiff at high conductivities.

In \cite{Komissarov2007}, the evolution system of equations was augmented by the addition of two scalar fields, $\phi$ and $\psi$, and their corresponding evolution equations. By doing this, the constraint for the divergence of the magnetic field, $\nabla\cdot\bm B = 0$, and its evolution equation, $\partial_t B + \nabla\times E = 0$, were replaced by the following two equations:
\begin{eqnarray}
	\partial_t \phi + \nabla\cdot\bm B = - \kappa \, \phi, \label{e:phi} \\
	\partial_t\bm B + \nabla\times \bm E + \nabla\phi=0.
\end{eqnarray}
Analogously, the constraint for the divergence of the electric field, $\nabla\cdot\bm E=q$, and its evolution equation, $-\partial_t \bm E+\nabla\times \bm B= \bm J$, were replaced by the following two equations:
\begin{eqnarray}
	\partial_t \psi + \nabla\cdot\bm E = q-\kappa \, \psi, \\
	-\partial_t\bm E + \nabla\times\bm B - \nabla\psi = \bm J, \label{e:E}
\end{eqnarray}
where $J^i = \sigma \, W [E^i + (v\times B)^i - (E_j v^j)\,v^i] + q\, v^i$, 
$W=(1-v^2)^{-1/2}$ is the Lorentz factor and $\sigma$ is the conductivity. In the definition of $J^i$ we are assuming Ohm's law for the current density. With these replacements and for positive $\kappa$, the potential constraint violations that may be generated numerically will decay exponentially and propagate at the speed of light. Moreover, after these replacements are applied, one only has to solve a system of evolution equations for the electromagnetic sector, formed by Eqs. (\ref{e:phi})--(\ref{e:E}) together with the evolution equation for the charge density $q$,
\begin{equation}
	\partial_t q + \nabla\cdot \bm J = 0.
\end{equation}

The electromagnetic sector has to be evolved together with the hydrodynamic equations, which can be written as
\begin{eqnarray}
	&&\partial_t D + \nabla\cdot (\rho \, W \, v) = 0,\label{eq:D} \\
	&&\partial_t\bm P + \nabla\cdot (- \bm E \, \bm E - \bm B \, \bm B + \rho \, h \, W^2 \, \bm v \,\bm v
    + [(E^2+B^2)/2 + p] \, \bm g) = 0, \label{eq:P}\\
	&&\partial_t e + \nabla\cdot (\bm E \times \bm B + \rho \, h \, W^2 \, \bm v) = 0,\label{eq:e}
\end{eqnarray}
where $D = \rho \, W$, $e = (E^2+B^2)/2+\rho \, h \, W^2 - p$, $h = 1 + \epsilon + p/\rho$ is the relativistic enthalpy per unit mass, $p$ is the thermodynamic pressure, $P^i = (\bm E\times \bm B)^i + \rho \, h \, W^2 \, v^i$ and $\bm g$ is the flat (Minkowski) space-time metric.

All the evolution equations can be written as a hyperbolic balance law.

In \cite{Komissarov2007} the same variable $\kappa$ was introduced in the replacement of both constraint equations. However, for the case of the general relativistic force-free electrodynamics, \cite{Mahlmann2021} found it more convenient to choose different values for $\kappa$ in each of the constraint equations, and the optimal values were actually very different (by approximately 3 orders of magnitude). In this manuscript we will follow the approach considered in \cite{Komissarov2007}, since our numerical experiments do not require different values for $\kappa$.

At this point, it is important to clarify the process of the recovery of the variables. On one hand, the set of physical variables, $\{\phi,B^i,\psi,E^i,q,\rho,\epsilon,v^i\}$, are called the primitive variables. On the other hand, the set of evolved variables, $\{\phi,B^i,\psi,E^i,q,D,P^i,e\}$, are called the conserved variables. The determinant of the matrix of change of variables is always different from zero, so a bijective relation between the primitive and conserved variables is always locally guaranteed to exist. Notice that the subset $\{\psi,E^i,\phi,B^i,q\}$, associated to the Maxwell equations, has elements which are both part of the primitive and the conserved variables. In one direction, the conserved variables can be obtained directly from the primitive ones from their definitions. In the other direction, this process is known as the recovery, and it can be quite difficult to obtain the explicit values of the primitive variables from the conserved ones in a general scenario. Although the set of quantities that we do evolve in time are the conserved variables, we are interested in the explicit values of the primitive and physical ones, and these values are also needed in order to compute the pressure $p$ appearing in the source terms.

Moreover, it is important to highlight that the conductivity $\sigma$ can be potentially large, so the source term of the evolution equation for the electric field, and therefore the whole system of equations, can be potentially stiff. The ideal regime is defined by the limit $\sigma \to \infty$, and in this case $E^i = - (\bm v\times \bm B)^i$. If not taken into account in the numerical resolution, the stiffness of the source term for $\sigma \gg 1$ can lead to the development of numerical instabilities. We can write the evolution equations, explicitly highlighting this stiff term, as follows:
\begin{eqnarray}
	\partial_t E^j &=& S_E^j - \sigma \, W [{\color{orange}{E}}^j 
    + (v\times{\color{orange}{B}})^j - (v_l{\color{orange}{E}}^l) v^j] 
    = \tilde{S}_E^j,\\
	\partial_t B^j &=& S_B^j, \\
	\partial_t Y &=& S_Y,
\label{e:structure}
\end{eqnarray}
where the reason for the terms in orange will be explained later and $Y$ denotes the rest of the evolved variables, $Y = \{\phi, \psi, q, D, e, P^i\}$. We will not include the set of variables $Y$ in the implicit terms, and this is the reason for considering this structure for the evolution equations. It is remarkable that in relativistic fluids the conductivity $\sigma$ always appears multiplied by the Lorentz factor $W$, so one could define an effective relativistic conductivity, $\bar{\sigma} = \sigma\,W$.

\section{Previous numerical approaches}
As mentioned in the previous section, the presence of stiff source terms requires a specific approach. One option is to implement an implicit treatment of the source term, or part of it. A hyperbolic equation with a relaxation term can be written as follows:
\begin{equation}
	\partial_t U = F(U) + \frac{1}{\epsilon} R(U).
\end{equation}
Here, $R(U)$ does not have derivatives with respect to the evolved variable $U$, and we have stiff source term in case $\epsilon \ll 1$.

In order to numerically solve the RRMHD evolution system of equations presented in the previous section, and taking into account the structure of a hyperbolic equation with a relaxation and stiff source term, some numerical methods have been used in the literature. For example, in \cite{Komissarov2007} the Strang-splitting method was applied. Also, the authors of reference \cite{Palenzuela2009} used IMEX Runge-Kutta methods (we refer to this reference for more details about the application of these methods). With this technique, these authors were able to successfully perform several simulations: the evolution of Alfv\'en waves, where high values for the waves amplitude and the conductivity were considered, and the results for the ideal case were properly recovered; the evolution of a self-similar current sheet; the evolution of shock tubes, where a broad range of different values for the conductivity was considered; or the evolution of a neutron star with magnetic field. In their approach, the implicit operator is also applied to the whole source term which contains the conductivity factor; in particular, this source term also contains the Lorentz factor and it is important to notice that it is defined in terms of primitive variables (specifically, in terms of the components of the velocity field), and therefore requires additional recoveries and iterative loops. This implies that the application of IMEX methods is very expensive computationally, and the nested iterative loops for the additional recoveries of primitive variables do not have any guarantee of convergence.

This motivates us to design an alternative approach, which is presented in the next section.

The authors of \cite{DumbserZanotti2009} use local space-time discontinuous Galerkin methods to deal with the stiffness of the source terms of the same RRMHD equations (again, we refer to this reference for more details), in the context of unstructured meshes in multiple space dimensions with an unified framework of one-step finite volume and discontinuous Galerkin schemes. A locally implicit scheme, explicit for the fluxes and implicit (but not minimally implicit, see next section) for the source, was used. These authors were able to successfully perform several simulations (some of them similar to the ones presented in \cite{Palenzuela2009}): the evolution of Alfv\'en waves, where high values for the waves amplitude and the conductivity were again considered, and the results for the ideal case were properly recovered; the evolution of a self-similar current sheet; the evolution of shock tubes, with different values for the conductivity; the resistive relativistic version of the MHD rotor problem; the cylindrical explosion problem; or the resistive relativistic analogous of the Orszag-Tang vortex problem.

It would be very interesting to compare these results, which include non-smooth data, with the ones obtained by using a minimally implicit (our new proposal, see next section) implementation for the source terms, but this is beyond the scope of this manuscript.

\section{Alternative approach: minimally-implicit Runge-Kutta methods}

In the structure introduced in Eq.~(\ref{e:structure}), the terms $S_Y$, $S_B^j$ and $S_E^j$ are evolved explicitly, and we only consider implicit evaluations of the electric and magnetic components, $E^j, B^j$, appearing in the source term for the evolution of the electric field multiplied by the conductivity $\sigma$ (orange terms in this system).

The general proposal in the case of a first-order method can be written as:
\begin{eqnarray}
	E^j|_{n+1} &=& E^j|_n + \Delta t \, S_E^j|_n - \Delta t \, \bar{\sigma}|_n \,
    [c_1 E^j|_n + (1-c_1) E^j|_{n+1} + c_2 (v\times B)^j|_n \nonumber \\
	&&+ (1-c_2) (v|_n\times B|_{n+1})^j - c_3 v^j|_n v_l|_n E^l|_n \nonumber \\ 
	&&- (1-c_3) v^j|_n v_l|_n E^l|_{n+1}],\\
	B^j|_{n+1} &=& B^j|_n + \Delta t \, S_B^j|_n,\\
	Y|_{n+1} &=& Y|_n + \Delta t \, S_Y|_n.
\end{eqnarray}
Since we want finite values for the computed quantities for very high values of the effective conductivity (i.e., $\bar{\sigma} \gg 1$), we will request $[1 - c_1 + v^2|_n (c_3 - 1)] \neq 0 \neq (1-c_1)$. Taking into account the wave-like behavior of the magnetic and electric fields, we use a first-order Partially Implicit RK (PIRK) method \cite{PIRKarxiv,PIRKproc}, which sets $c_2=0$.

We perform a linear stability analysis of the evolution system in the case of infinite conductivity. We can choose $c_3=1$ keeping numerical stability of the numerically evolved system, slightly simplifying the proposed method. Moreover, in the ideal limit $\sigma\to\infty$, the electric field $E^i$ is no longer an independent quantity, since $E^i=(v\times B)^i$; this means that, independently of the velocity field $v^i$, we should have a zero eigenvalue of multiplicity at least 3 and we need to impose $c_1=0$. With these choices, the other eigenvalue of the system is bounded by 1 in absolute value, independently of the velocity field $v^i$ value considered. In addition, $[1 - c_1 + v^2|_n (c_3 - 1)] = (1-c_1) = 1 \neq 0$ as requested previously, and the ideal limit is recovered for $\Delta t \to 0$.

The numerical integration of the evolution equation for $E^i$ can then be written as:
\begin{equation}
	E^i|_{n+1} = E^i|_n + \frac{\Delta t}{1+\Delta t\,\bar{\sigma}|_n} 
    [S^i_E|_n + \bar{\sigma}|_n\,E^l|_n (v^i|_n v_l|_n-\delta^i_l) 
    - \bar{\sigma}|_n (v|_n\times B|_{n+1})^i].
\end{equation}
This scheme can be viewed as an explicit scheme for the evolution of the electric field when an effective time-step, $\displaystyle \frac{\Delta t}{1+\Delta t\,\bar{\sigma}|_n}$, is considered. This effective time step is of the order of $\Delta t$, for sufficiently small values of this quantity. In some sense, we are implementing a numerical first-order explicit method for the modified evolution equation $\displaystyle \partial_t E^j = \frac{1}{1+\Delta t\,\bar{\sigma}|_n} \tilde{S}_E^j$, where the modification is of order $\Delta t$ and thus we recover the original evolution equation for the electric field in the limit $\Delta t \to 0$. Adapting explicit schemes to this method is direct.

The general proposal for a second-order two-stages method can be written as:
\begin{eqnarray}
	E^j|_{(1)} &=& E^j|_n + \Delta t \, S_E^j|_n - \Delta t \, \bar{\sigma}|_n \,
    [c_1 E^j|_n + (1-c_1) E^j|_{(1)} + c_2 (v\times B)^j|_n \nonumber \\
	&&+ (1-c_2) (v|_n\times B|_{(1)})^j - c_3 v^j|_n v_l|_n E^l|_n \nonumber \\ 
	&&- (1-c_3) v^j|_n v_l|_n E^l|_{(1)}],\\
	B^j|_{(1)} &=& B^j|_n + \Delta t \, S_B^j|_n,\\
	Y|_{(1)} &=& Y|_n + \Delta t \, S_Y|_n.
\end{eqnarray}
\begin{eqnarray}
	E^j|_{n+1} &=& \frac{1}{2} [E^j|_n + E^j|_{(1)} + \Delta t \, S_E^j|_{(1)}]
    - \Delta t \, \bar{\sigma}|_{(1)} \left[
    \frac{(1-c_1)}{2} E^j|_n + c_4 E^j|_{(1)} \right. \nonumber \\ 
	&&+ \left(\frac{c_1}{2}-c_4\right) E^j|_{n+1} 
    + \frac{1-c_2}{2} (v|_{(1)} \times B|_n)^j + c_5 (v\times B)^j|_{(1)} \nonumber \\
	&&+ \left(\frac{c_2}{2}-c_5\right) (v|_{(1)}\times B|_{n+1})^j \nonumber \\
	&&\left.+ v^j|_{(1)} v_l|_{(1)} \left( \frac{(1-c_3)}{2} E^l|_n + c_6 E^l|_{(1)}
    + \left(\frac{c_3}{2}-c_6\right)E^l|_{n+1}\right) \right],\\
	B^j|_{n+1} &=& \frac{1}{2} [B^j|_n + B^j|_{(1)} + \Delta t \, S_B^j|_{(1)}],\\
	Y|_{n+1} &=& \frac{1}{2} [Y|_n + Y|_{(1)} + \Delta t \, S_Y|_{(1)}].
\end{eqnarray}

Since we want finite values for the computed quantities for very high values of the effective conductivity (i.e., $\bar{\sigma}\gg 1$), we will request $[1 - c_1 + v^2|_n (c_3 - 1)] \neq 0 \neq (1-c_1)$ and $[c_1/2-c_4-v^2|_{(1)} (c_3/2-c_6)] \neq 0 \neq (c_1/2-c_4)$. Using a second-order PIRK method for the wave-like behavior of the electric and magnetic fields sets $c_2=1-\frac{\sqrt{2}}{2}$, $c_5=(\sqrt{2}-1)/2$.

As in the first-order method, we perform a linear stability analysis of the evolution system in the case of infinite conductivity. We can choose $c_3=1$ and $c_6=1/2$, keeping numerical stability of the numerically evolved system, slightly simplifying the proposed method. Moreover, in the ideal limit $\sigma\to\infty$, due to the same reason as in the first-order method, we need to impose $c_4=\frac{(1-c_1)^2}{2c_1}$, $c_1 \neq 0$, so one eigenvalue of multiplicity at least 3 is set to zero. With these choices, the other eigenvalue of the system is bounded by 1 in absolute value, independently of the velocity field $v^i$ value considered, if $c_1<0$; actually the expression for this eigenvalue achieves its minimum in absolute value with respect to the remaining coefficient $c_1$ for $c_1=-1/\sqrt{2}$. We will choose this value for the $c_1$ coefficient. Finally, $[1 - c_1 + v^2|_n (c_3 - 1)] = (1-c_1) = (1 + 1/\sqrt{2}) \neq 0$, $[c_1/2-c_4-v^2|_{(1)} (c_3/2-c_6)] = (c_1/2-c_4) = (1 + \sqrt{2}/2) \neq 0$, and the ideal limit is recovered for $\Delta t \to 0$.

The two-stages of the numerical integration of the evolution equation for $E^i$ can then be written as follows:
\begin{eqnarray}
	E^j|_{(1)} &=& E^j|_n + \Delta t \, S_E^j|_n - \Delta t\,\bar{\sigma}|_n \,
    \left[-\frac{1}{\sqrt{2}} E^j|_n + (1+1/\sqrt{2}) E^j|_{(1)} \right. \nonumber \\
	&&\left. + (1-1/\sqrt{2}) (v\times B)^j|_n + \frac{1}{\sqrt{2}} (v|_n\times B|_{(1)})^j 
    - v^j|_n v_l|_n E^l|_n \right],
\label{e:MIRK2-1}
\end{eqnarray}
\begin{eqnarray}
	E^j|_{n+1} &=& \frac{1}{2} [E^j|_n + E^j|_{(1)} + \Delta t \, S_E^j|_{(1)}]
    - \Delta t \, \bar{\sigma}|_{(1)} \left[ \frac{(1+1/\sqrt{2})}{2} E^j|_n \right. \nonumber \\ 
	&&- \frac{\sqrt{2}(1+\sqrt{2})^2}{4} E^j|_{(1)} + (1+\sqrt{2}/2) E^j|_{n+1}
    + \frac{1}{2\sqrt{2}} (v|_{(1)} \times B|_n)^j \nonumber \\
	&&+ \frac{(\sqrt{2}-1)}{2} (v\times B)^j|_{(1)}
    + (1-3\sqrt{2}/4) (v|_{(1)}\times B|_{n+1})^j \nonumber \\
	&&\left.+ \frac{1}{2} v^j|_{(1)} v_l|_{(1)} E^l|_{(1)} \right].
\label{e:MIRK2-2}
\end{eqnarray}

Eq. (\ref{e:MIRK2-1}) can be rewritten as:
\begin{eqnarray}
	E^j|_{(1)} &=& E^j|_n + \frac{\Delta t}{1+\Delta t \, \bar{\sigma}|_n (1+1/\sqrt{2})} 
    \left[ S_E^j|_n + \bar{\sigma}|_n E^l|_n (v^j|_n v_l|_n-\delta^j_l) \right. \nonumber \\
	&&\left. - \bar{\sigma}|_n (1-1/\sqrt{2}) (v\times B)^j|_n 
    - \frac{\bar{\sigma}|_n}{\sqrt{2}} (v|_n\times B|_{(1)})^j \right].
\end{eqnarray}
Eq. (\ref{e:MIRK2-2}) can be rewritten as:
\begin{eqnarray}
	E^j|_{n+1} &=& \frac{1}{2} (E^j|_n + E^j|_{(1)})
    + \frac{\Delta t}{1+\Delta t \, \bar{\sigma}|_{(1)} (1+1/\sqrt{2})} \left[
    \frac{1}{2} S_E^j|_{(1)} \right. \nonumber \\
	&&- \bar{\sigma}|_{(1)} (1+1/\sqrt{2}) E^j|_n
    + \bar{\sigma}|_{(1)} \frac{(1+\sqrt{2})}{2} E^j|_{(1)} \nonumber \\
	&&- \bar{\sigma}|_{(1)} \frac{\sqrt{2}}{4} (v|_{(1)} \times B|_n)^j 
    - \bar{\sigma}|_{(1)} \frac{(\sqrt{2}-1)}{2} (v\times B)^j|_{(1)} \nonumber \\
	&&\left. - \bar{\sigma}|_{(1)} (1-3\sqrt{2}/4) (v|_{(1)}\times B|_{n+1})^j
    - \bar{\sigma}|_{(1)} \frac{1}{2} v^j|_{(1)} v_l|_{(1)} E^l|_{(1)} \right]. \;\;\;\;\;\;
\end{eqnarray}
Here, again, an effective time step, $\displaystyle \frac{\Delta t}{1+\Delta t \, \bar{\sigma} (1+1/\sqrt{2})}$, appears. For the first stage, $\bar{\sigma}$ is evaluated in the previous time-step, $\bar{\sigma}|_n$. For the second stage, $\bar{\sigma}$ is evaluated in the first stage, $\bar{\sigma}|_{(1)}$.

\section{Numerical simulations}

We use Cartesian coordinates, equally spaced numerical grid and centered finite differences of second order for the discretization of the spatial derivatives. We present two different numerical tests, namely the self-similar current sheet test and the large amplitude Alfv\'en wave test; both tests deal with smooth initial data and smooth data during the evolution, and are also discussed in detail in references \cite{Palenzuela2009,DumbserZanotti2009} to check the success of their numerical codes when dealing with high values for the conductivity.\\

\begin{figure}[h]
\centering
	\subfloat[$B^y$]{
  \label{fig:By1}
  \includegraphics[width=0.49\textwidth]{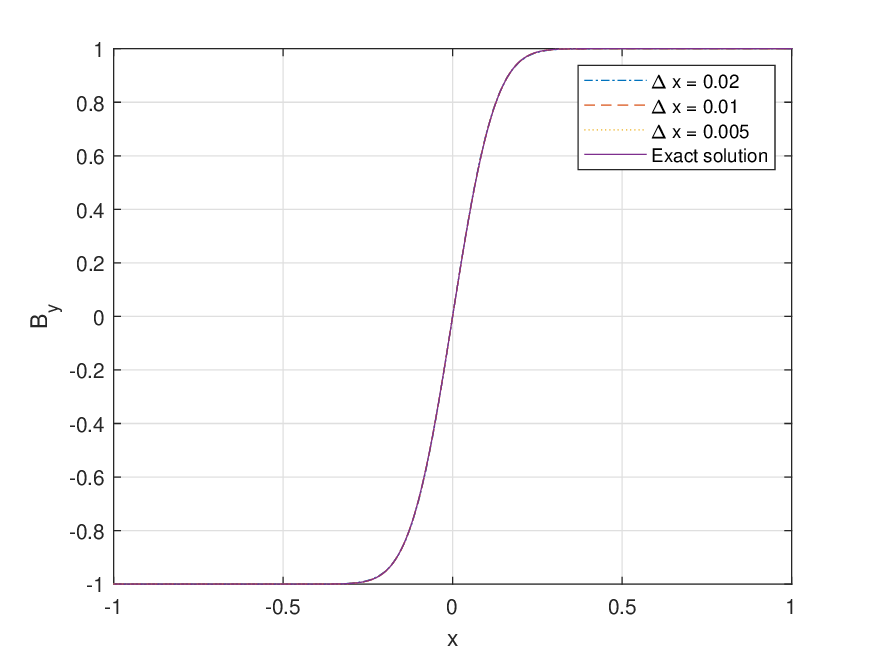}}
  \subfloat[$E^z$]{
  \label{fig:Ez1}
  \includegraphics[width=0.49\textwidth]{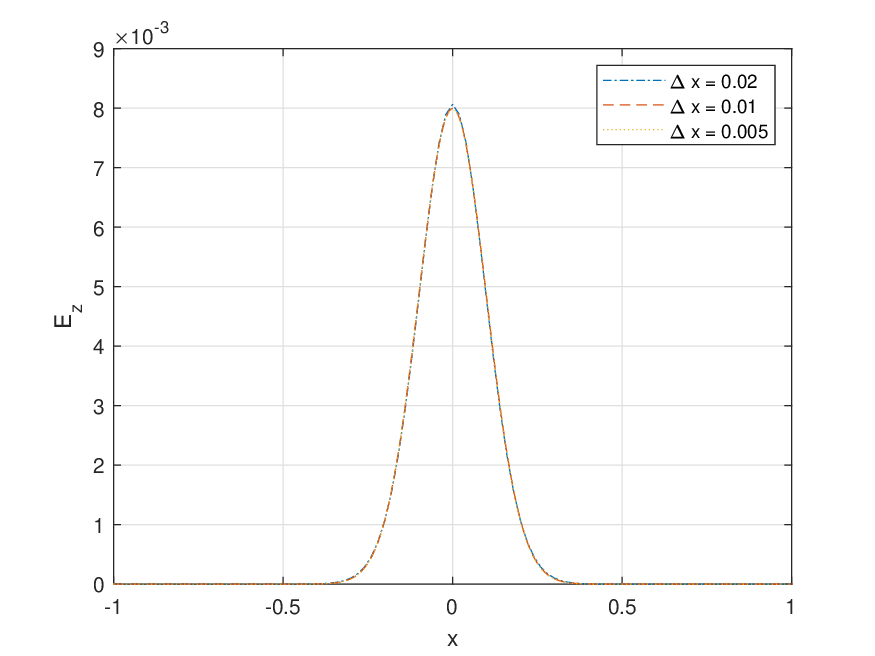}}
 \caption{Self similar current sheet test. Numerical values for $B^y$ and $E^z$ at $t=5$, for three different spatial resolutions, when a first-order MIRK method is used. $v_x=0$, $\sigma=10^3$ and CFL=0.8. The exact solution is also included.}
\end{figure}

In general, the evolution of the magnetic and electric fields are considered, and the charge is computed from its definition as divergence of the electric field. The zero divergence of the magnetic field is conserved through the evolution due to the particular configuration of initial data. We present numerical results using the first and second-order MIRK methods.

\subsection{Self similar current sheet}
This is a simple test in 1D. During the evolution, the only non-zero components of the electric and magnetic fields are $B^y(x,t)$ and $E^z(x,t)$. Vacuum is considered, so we do not evolve the hydrodynamic sector. We keep both the velocity components and the conductivity as constant values. Unless otherwise stated, a Courant-Friedrichs-Lewy (CFL) value of 0.8 is used. The set-up for the initial data is: $\phi=0$, $\bm v=(v^x,0,0)$, $\bm E=(0,0,0)$, $\bm B=(0,B^y(x,t=1),0)$, being $B^y(x,t=1)= \mbox{erf} ( x\sqrt{\sigma}/2)$. The exact solution of this problem is
\begin{equation}
    B_e^y(x,t) = \mbox{erf}\left(\frac{x}{2}\sqrt{\frac{\sigma}{t}}\right).
\end{equation}
We consider $t=1$ for the initial data to avoid singular values of the exact solution. We consider $x\in[-1,1]$. At the spatial boundaries $x=-1$ and $x=1$, we make use of ghost cells, where the evolved variables are set equal to the the values from the adjacent cells inside our numerical domain. We explore two illustrative examples among the possibilities for this very simple case.

\begin{figure}[h]
\centering
	\subfloat[$B^y$]{
  \label{fig:5}
  \includegraphics[width=0.49\textwidth]{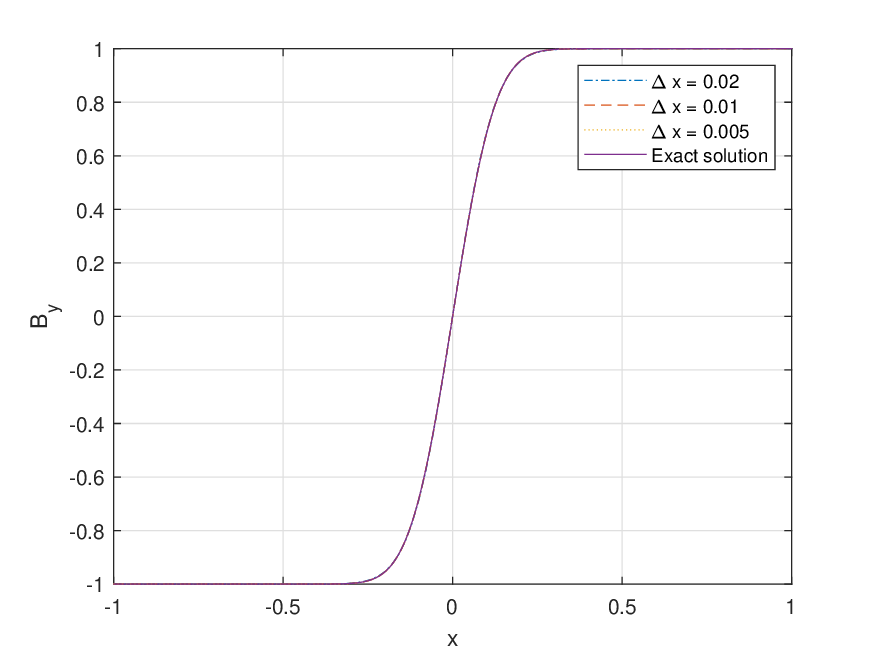}}
  \subfloat[$E^z$]{
  \label{fig:6}
  \includegraphics[width=0.49\textwidth]{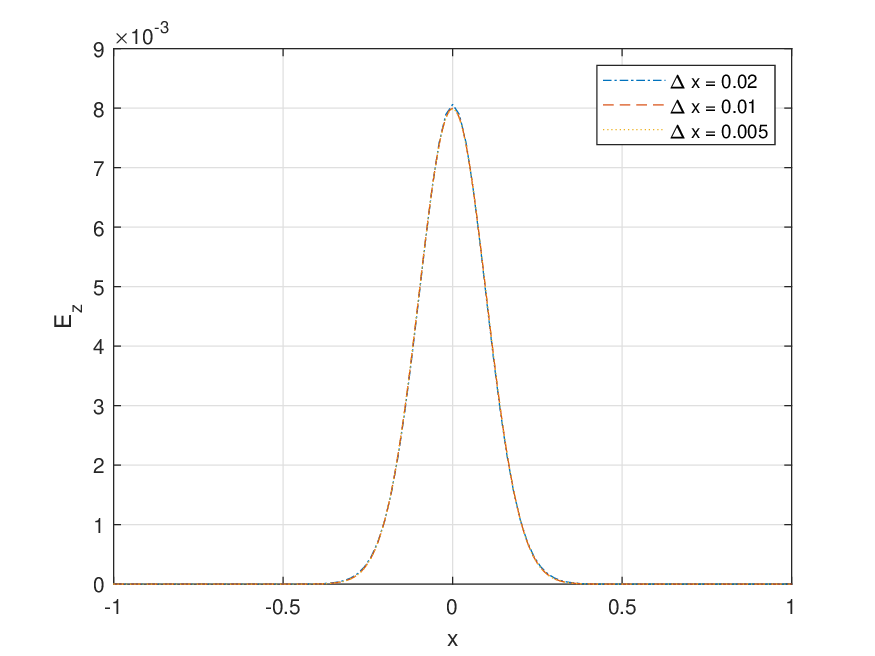}}
 \caption{Self similar current sheet test. Numerical values for $B^y$ and $E^z$ at $t=5$, for three different spatial resolutions, when a second-order MIRK method is used. $v_x=0$, $\sigma=10^3$ and CFL=0.8. The exact solution is also included.}
\end{figure}
\begin{figure}[h]
\centering
  \subfloat[$B^y$. First-order method.]{
  \label{fig:By1z}
  \includegraphics[width=0.49\textwidth]{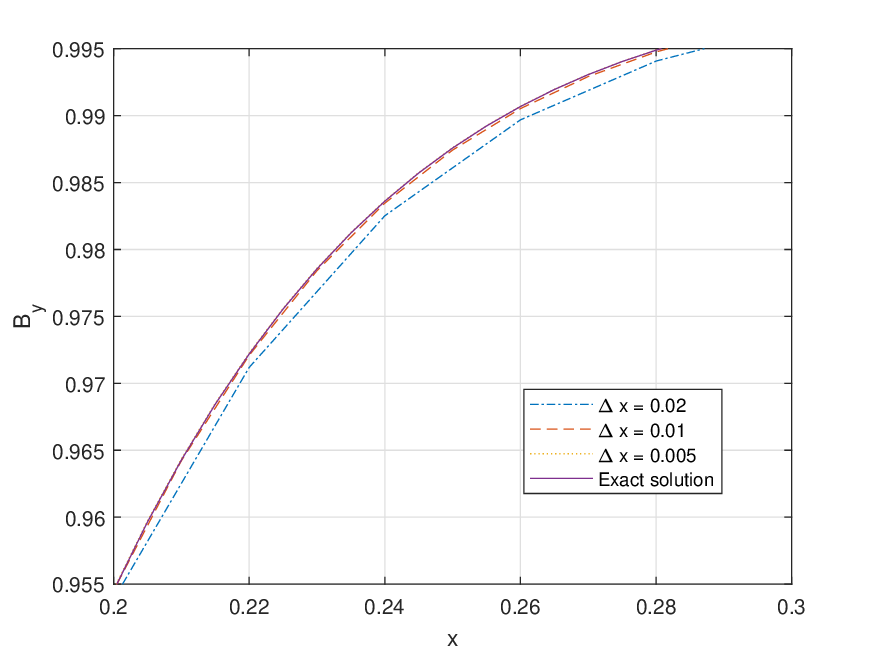}}
  \subfloat[$B^y$. Second-order method.]{
  \label{fig:By2z}
  \includegraphics[width=0.49\textwidth]{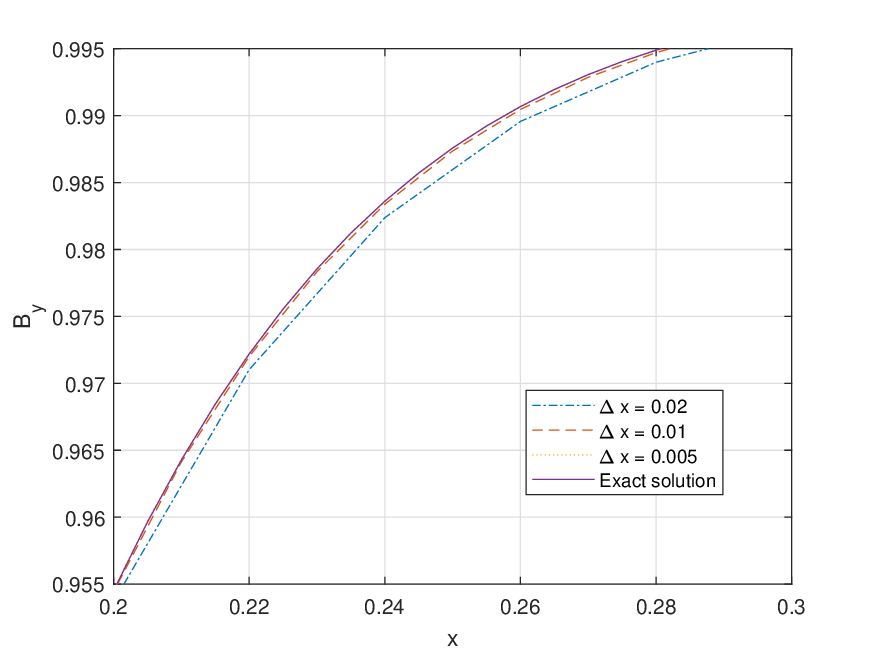}}\\
  \subfloat[$E^z$. First-order method.]{
  \label{fig:Ez1z}
  \includegraphics[width=0.49\textwidth]{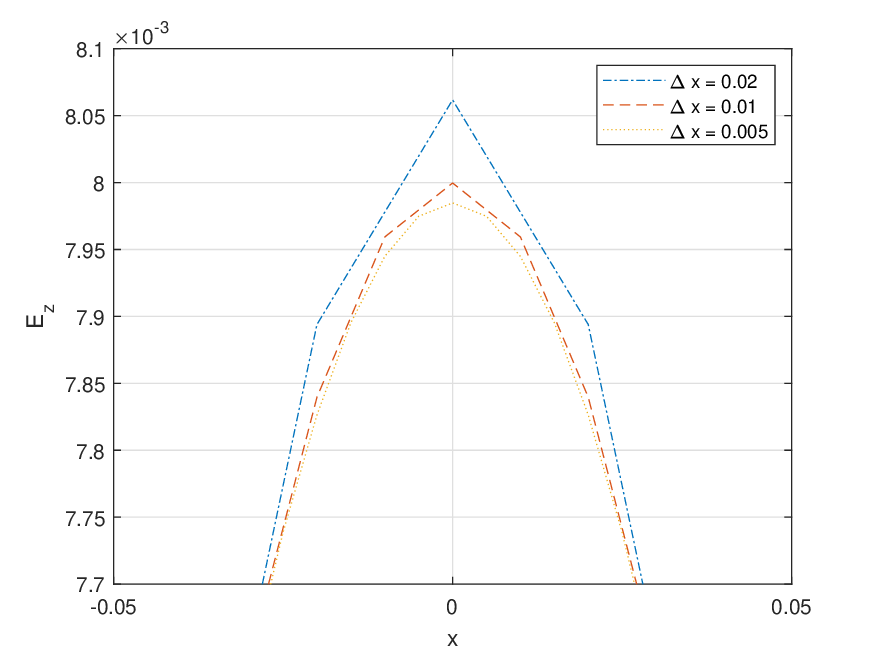}}
  \subfloat[$E^z$. Second-order method.]{
  \label{fig:Ez2z}
  \includegraphics[width=0.49\textwidth]{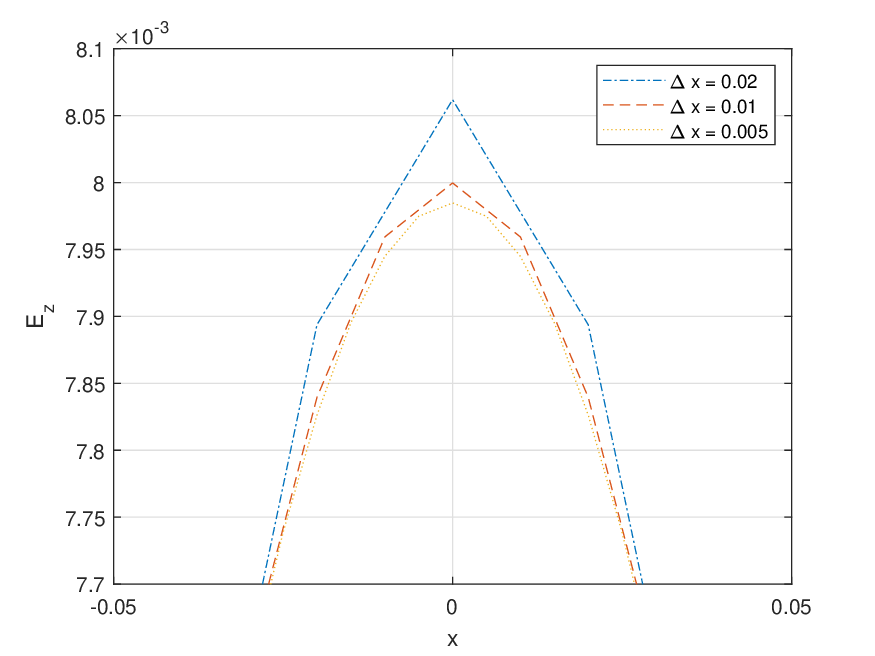}}
 \caption{Self similar current sheet test. Zoom on the numerical values for $B^y$ and $E^z$ at $t=5$, for three different spatial resolutions, when a first and second-order MIRK methods are used. $v_x=0$, $\sigma=10^3$ and CFL=0.8.}
\label{fig:zoom}
\end{figure}
\begin{table}[htbp!]
\begin{center}
\begin{tabular}{|c||c|c|c|}
\hline
 $\Delta x$ & $0.04$ & $0.02$ & $0.01$ \\ \hline \hline
  $p$ for the 1st-order MIRK method & 2.1648142 & 2.0709731 & 2.0383649 \\\hline
  $p$ for the 2nd-order MIRK method &  2.1522280 & 2.0501579 & 2.0197818
  \\\hline
\end{tabular}
\end{center}
\caption{Estimated convergence orders for the first and second-order MIRK methods applied to the self similar current sheet test at $t=5$, with $\sigma = 10^3$ and CFL = $0.5$, and resolutions $\Delta x=0.04/2^k$, $k=0,1,2,3$, according to the formula (\ref{eq:ord}).}
\label{tab:ordercur}
\end{table}

On one hand, we consider $v^x=0$ and $\sigma=10^3$. The results for $B^y$ and $E^z$ for three different spatial resolutions (namely, $\Delta x= 0.02$, $0.01$ and $0.005$) when the first-order MIRK method is used, together with the exact solution, are displayed in Figures \ref{fig:By1} and \ref{fig:Ez1} at $t=5$. The same data are displayed in Figures \ref{fig:5} and \ref{fig:6} when a second-order MIRK method is used.

In all cases, the numerical values are on top of the exact solution. We get convergence of the evolved variables, as can be better appreciated in a zoom of the previous Figures, displayed in Figure \ref{fig:zoom}. No significant differences between the results of first and second-order MIRK methods are found. Second-order convergence is obtained when both methods are applied using the $L_2$ norm of the error between the numerical and the analytical (available for this test) solutions, always using points from the coarsest grid. Specifically, we are considering the following formula:
\begin{equation}
    p\approx \log_2\left(\frac{\varepsilon(\Delta x)}{\varepsilon(\Delta x/2)}\right),
    \label{eq:ord}
\end{equation}
where $p$ is the estimate of the order of convergence and $\varepsilon(\Delta x)$ is the $L_2$ norm of the error of a numerical solution with respect to the analytical one for a resolution $\Delta x$. In Table \ref{tab:ordercur} we show the estimated order of convergence for several resolutions. We get second-order of convergence for both first and second-order MIRK methods. It is remarkable that second-order is achieved also for the first-order MIRK method. This is, most probably, due to the fact that this is indeed a very simple test where the solution is symmetric with respect to $x=0$.

\begin{figure}[t]
\centering
	\subfloat[First-order MIRK method.]{\label{fig:CFL11}
    \includegraphics[width=0.49\textwidth]{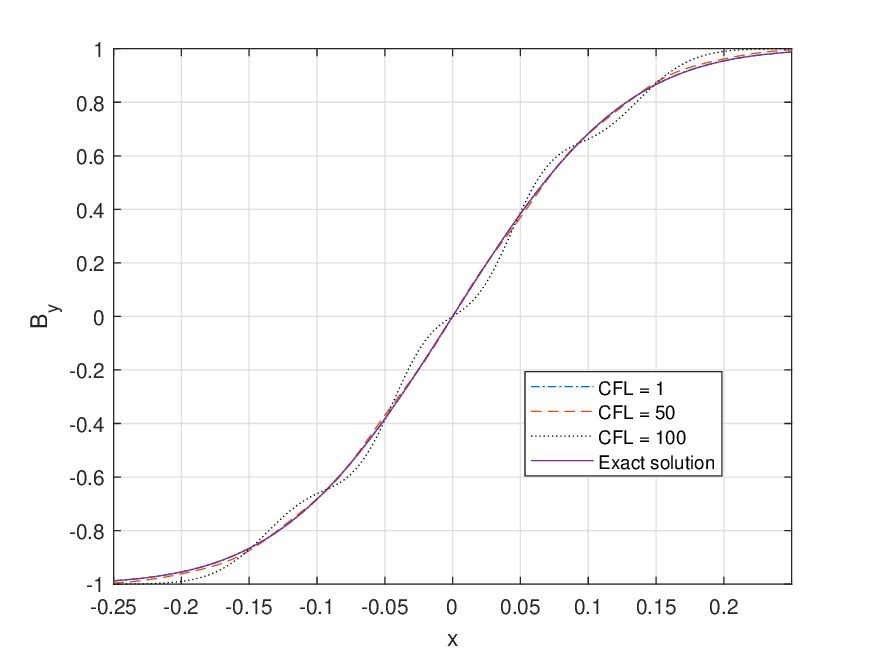}}
    \subfloat[Second-order MIRK method.]{\label{fig:CFL12}
    \includegraphics[width=0.49\textwidth]{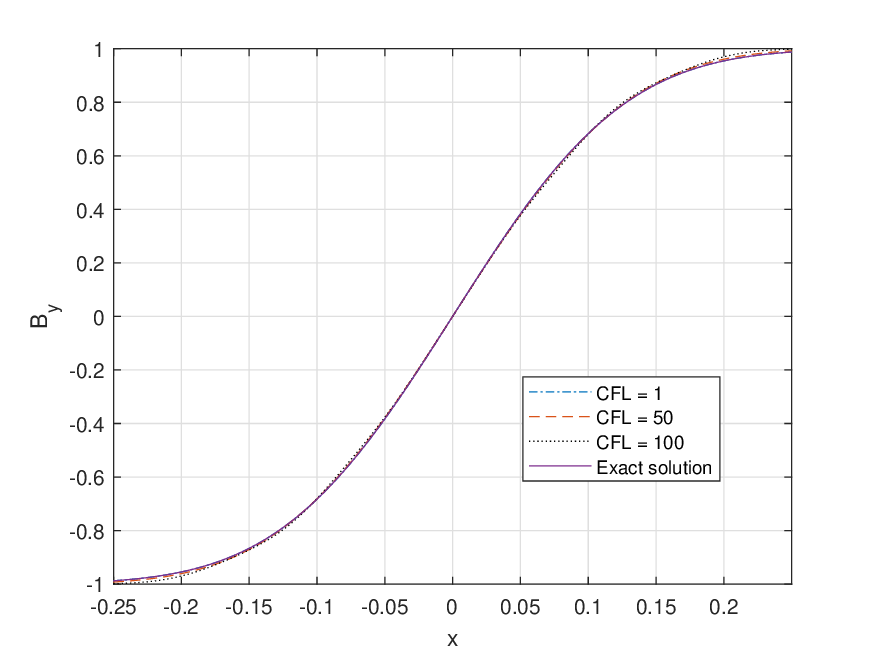}} \\
    \caption{Self similar current sheet test. Numerical results for several CFL values when first and second-order MIRK methods are used at $t=5$. $v_x=0$, $\sigma=10^3$ and $\Delta x=0.005$.}
\label{fig:CFL2}
\end{figure}

We explore now several CFL values, even higher than 1, to check the stability of the simulations, using first and second-order MIRK methods. We choose CFL values of 1, 50 and 100. We can observe in Figures \ref{fig:CFL11} and \ref{fig:CFL12} the appearance of numerical oscillations for very high CFL values. Increasing the order of the method improves this behaviour, making the numerical simulations for a CFL value of 100 stable at $t=5$. The appearance of these oscillations is not due to the use of MIRK methods; for large CFL values and using IMEX methods, we will find a similar behaviour \cite{privatePalenzuela}. The reason of these oscillations is the consequence of using a larger CFL than allowed for the source terms included in the purely explicit part in the MIRK methods ($S^j_E, S^j_B, S_Y$), producing this oscillatory behaviour.

\begin{figure}[h]
\centering
	\subfloat[$B^y$]{
  \label{fig:By1exp03}
  \includegraphics[width=0.49\textwidth]{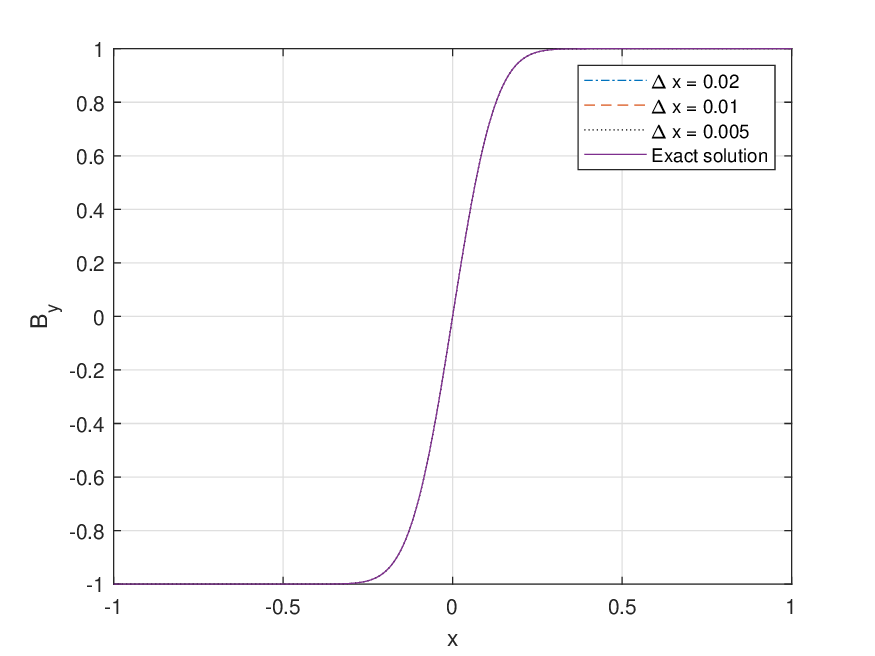}}
  \subfloat[$E^z$]{
  \label{fig:Ez1exp03}
  \includegraphics[width=0.49\textwidth]{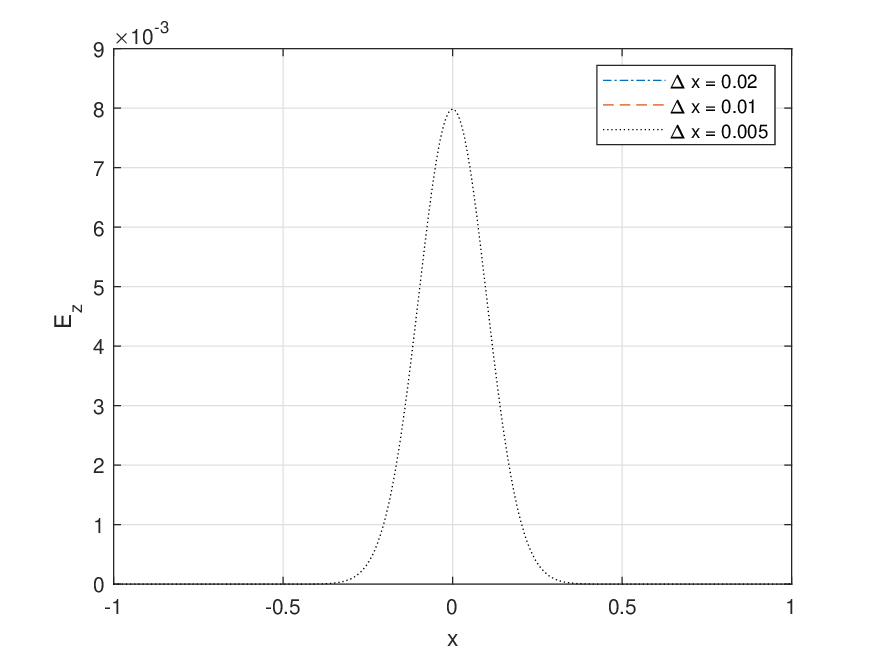}}
 \caption{Self similar current sheet test. Numerical values for $B^y$ and $E^z$ at $t=5$, for three different spatial resolutions, when a first-order explicit method is used. $v_x=0$, $\sigma=10^3$ and CFL=0.3. The exact solution is also included.}
\end{figure}
\begin{figure}[htbp!]
\centering
	\subfloat[$B^y$]{
  \label{fig:Byvexp03}
  \includegraphics[width=0.49\textwidth]{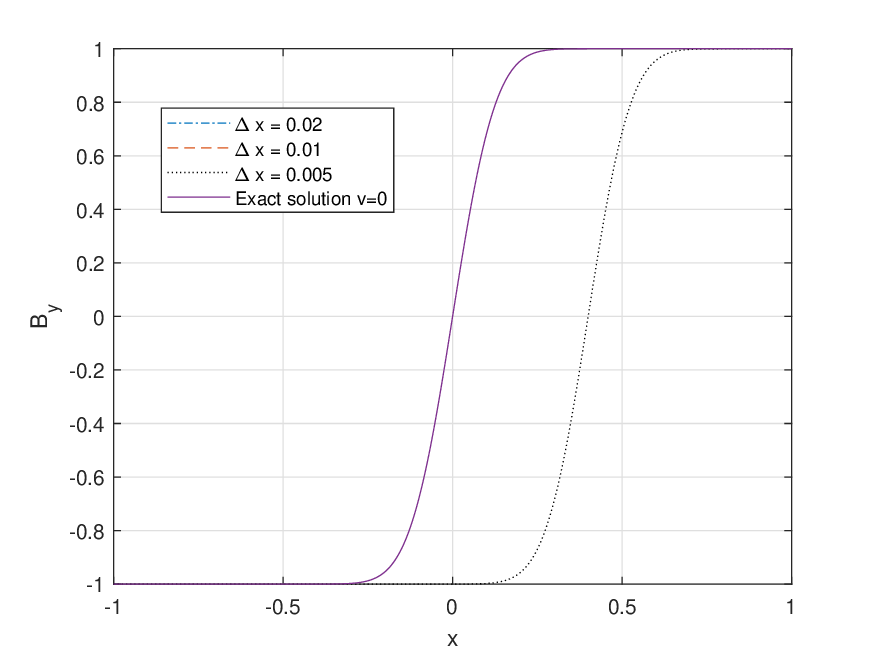}}
  \subfloat[$E^z$]{
  \label{fig:Ezvexp03}
  \includegraphics[width=0.49\textwidth]{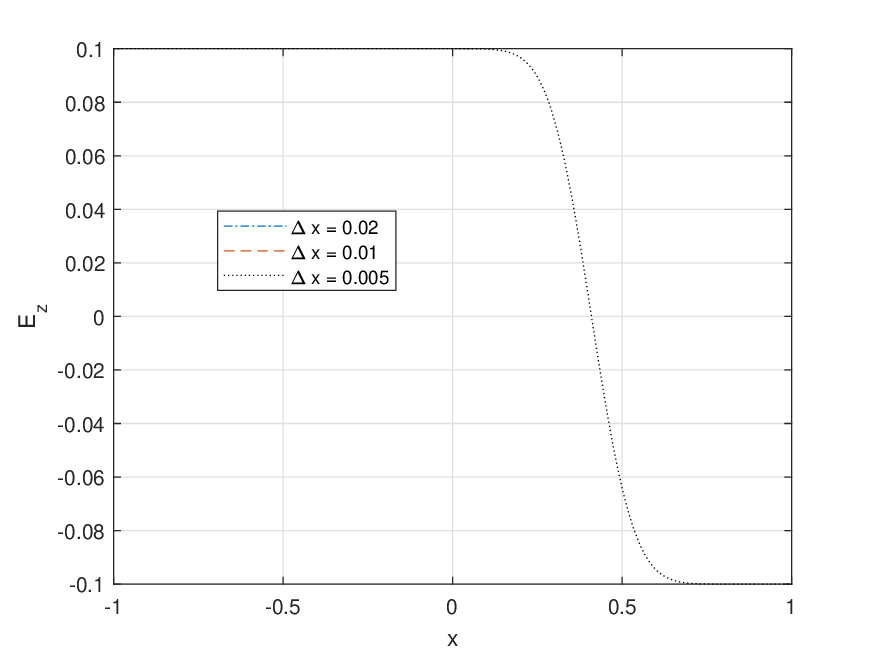}}
 \caption{Self similar current sheet test. Numerical values for $B^y$ and $E^z$ at $t=5$, for three different spatial resolutions, when a first-order explicit method is used. $v_x=0.1$, $\sigma=10^3$ and CFL=0.3. The exact solution is also included.}
\end{figure}

In the case of the first-order (pure) explicit method, $\Delta x=0.02$ and CFL=0.8, numerical instabilities develop very quickly, the electric and magnetic field components achieving values of $10^{290}$ at $t=5$. In order to get stable and accurate results, we need to consider $\Delta x = 0.005$ and CFL=0.3, as shown in Figures \ref{fig:By1exp03} and \ref{fig:Ez1exp03}. We find an analogous behaviour for the second-order explicit method.

On the other hand, we consider $v_x=0.1$ and $\sigma=10^3$. In the case of the first-order explicit method, $\Delta x = 0.02$ and CFL=0.8, the electric and magnetic field components develop again numerical instabilities very quickly, achieving values of order $10^{291}$ or higher at $t=5$. As previously for $v^x=0$, setting $\Delta x= 0.005$ and CFL=0.3, we obtain good numerical results, as shown in Figures \ref{fig:Byvexp03} and \ref{fig:Ezvexp03}. We find an analogous behaviour for the second-order explicit method.

\begin{figure}[h]
\centering
	\subfloat[$B^y$]{
  \label{fig:Byv}
  \includegraphics[width=0.49\textwidth]{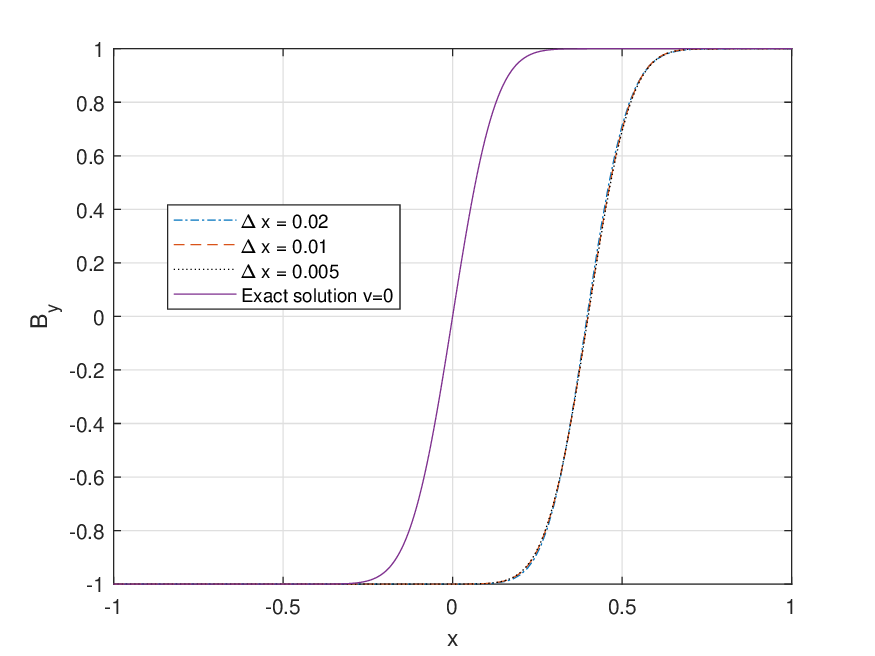}}
  \subfloat[$E^z$]{
  \label{fig:Ezv}
  \includegraphics[width=0.49\textwidth]{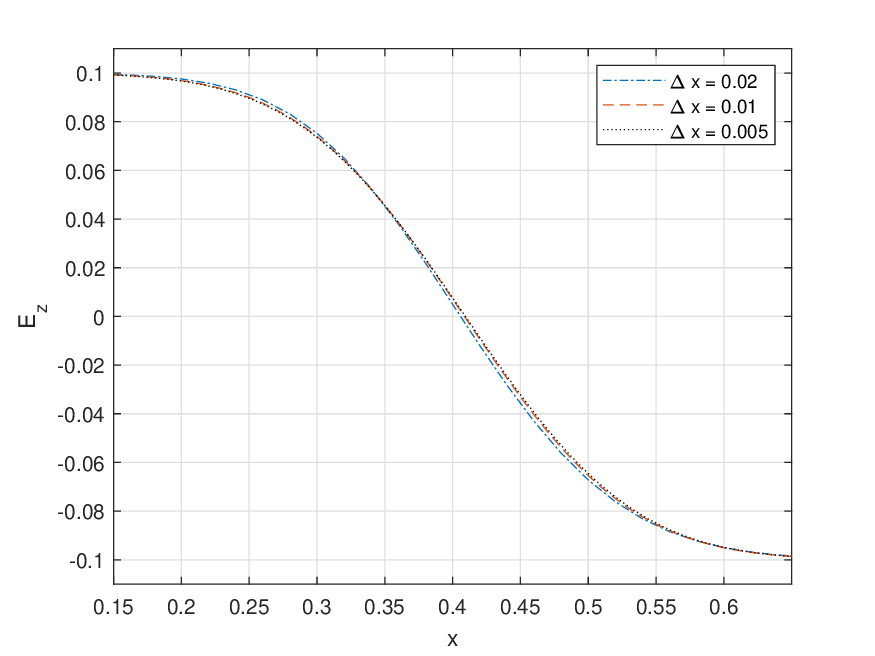}}
 \caption{Self similar current sheet test. Numerical values for $B^y$ and $E^z$ at $t=5$, for three different spatial resolutions, when first and second-order MIRK methods are used. $v_x=0.1$, $\sigma=10^3$ and CFL=0.8. The exact solution is also included.}
\end{figure}
\begin{figure}[htbp!]
\centering
	\subfloat[First-order method]{\label{fig:CFL11v}
  \includegraphics[width=0.49\textwidth]{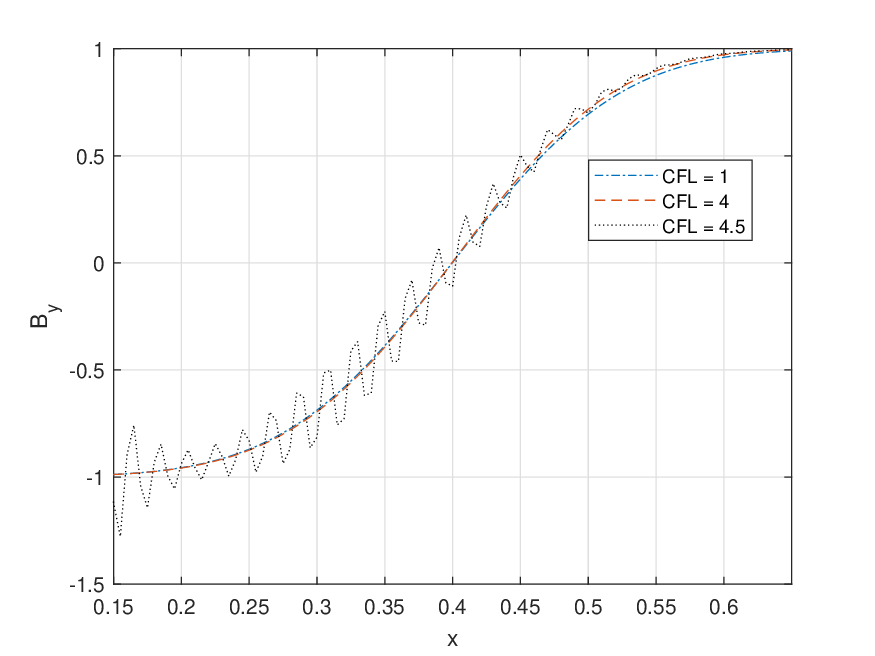}}
  \subfloat[Second-order method]{\label{fig:CFL12v}
  \includegraphics[width=0.49\textwidth]{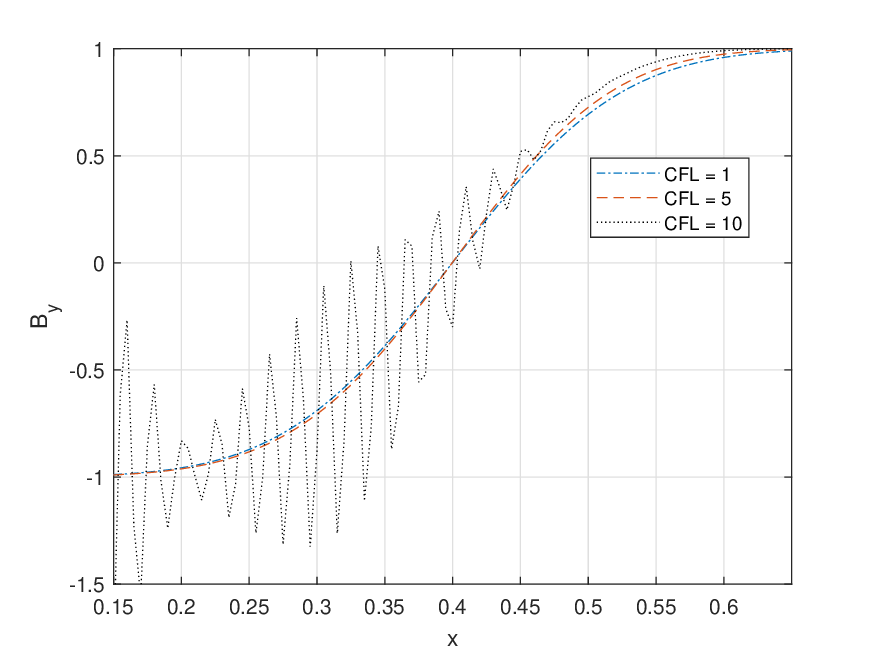}} \\
 \caption{Self similar current sheet test. Numerical results for several CFL values when first and second-order MIRK methods are used at $t=5$. $v_x=0.1$, $\sigma=10^3$ and $\Delta x=0.005$.}
\label{fig:CFL2}
\end{figure}

Instead, when the first-order MIRK method is used, the $B^y$ and $E^z$ profiles, shown in Figures \ref{fig:Byv} and \ref{fig:Ezv}, respectively, have the expected behavior for all resolutions: initial profiles are shifted to the right and slightly smoothed with time. All the profiles lie on top of the exact solution.

We also explore several CFL values. We see in Figures \ref{fig:CFL11v} and \ref{fig:CFL12v} that much lower CFL values are allowed; the oscillations begin to appear with CFL=4.5 for the first-order method and with CFL=10 for the second-order one. Again, the reason of these oscillations is associated to the stability of the terms included in the purely explicit part in the MIRK methods ($S^j_E, S^j_B, S_Y$).

\subsection{CP Alfven waves}

In this second test we simulate the Circular Polarized (CP) Alfv\'en Waves in 1D. Here the hydrodynamic Eqs. (\ref{eq:D})--(\ref{eq:e}) need to be solved, as well as the Maxwell Equations. The set-up for the initial data for the electromagnetic field is
\begin{equation}
\bm B (x, 0) = B_0 \, (1, \cos(k\,x), \sin(k\,x) ),
\end{equation}
with $k=2\pi$ and $B_0=1.1547$, and
\begin{equation}
\bm E (x, 0) = -\bm v(x,0) \times \bm B(x,0),
\end{equation}
with $\bm v(x,0) = \frac{v_A}{B_0} (0, B^y(x,0), B^z(x,0))$ and $v_A=0.423695$. Moreover, we consider $\rho(x,0)=p(x,0)= 1$ and the initial values for conserved hydrodynamic variables can be derived from them. We consider a perfect fluid model and an ideal fluid equation of state with $\Gamma = 4/3$, being $\Gamma$ the adiabatic index. $x\in[0,1]$ defines our numerical domain (one spatial period) and we impose periodic boundary conditions. Since we want to test if we are able to recover the ideal MHD limit, we consider $\sigma=10^8$ in all the simulations presented in the manuscript; the numerical solution should be very close to the analytical solution in the ideal limit.

Primitive variables must be computed on each time stage of each iteration; we apply the recovery procedure used in \cite{DumbserZanotti2009}. First, a quartic equation for the Lorentz factor $W$ can be derived, with coefficients defined in terms of conserved variables:
\begin{equation}
    A_4 W^4 + A_3 W^3 + A_2 W^2 + A_1 W + A_0 =0,
    \label{eq:W}
\end{equation}
where $A_0 = \gamma_1^2 \, (C_1+D^2)$, $A_1 = -2\gamma_1 \, C_2 \, D$, $A_2 = C_2^2 - 2\gamma_1 \, C_1 - \gamma_1^2 \, D$, $A_3 = 2\gamma_1 \, C_2 \, D$, $A_4 = C_1 - C_2^2$, $C_1 = |\bm S-\bm E\times \bm B|^2$, $C_2 = e - (E^2+B^2)/2$ and $\gamma_1=(\Gamma-1)/\Gamma$. We use the bisection method to get the solution of Eq. (\ref{eq:W}) with machine precision. Afterwards, we can compute the remaining primitive variables as follows:
\begin{eqnarray}
&& \rho = \frac{D}{W}, \\
&& h = \frac{e-\frac{1}{2}(E^2+B^2)-\gamma_1\frac{D}{W}}{W^2-\gamma_1},\\
&& p= \gamma_1(h-\rho),\\
&& \bm v = \frac{\bm P - \bm E\times\bm B}{e-\frac{1}{2}(E^2+B^2)+p}.
\end{eqnarray}

We consider CFL=0.3. A Kreiss-Oliger term of the form 
\begin{equation}
-\frac{\epsilon}{16}(\Delta x)^3\partial^4_x Y
\label{eq:KO}
\end{equation}
has been included in the hydrodynamic sector, in order to guarantee stability by adding a controlled amount of artificial dissipation, with $\epsilon = 0.01$.

\begin{figure}[t]
\centering
	\subfloat[$B^y$]{
  \label{fig:By}
  \includegraphics[width=0.49\textwidth]{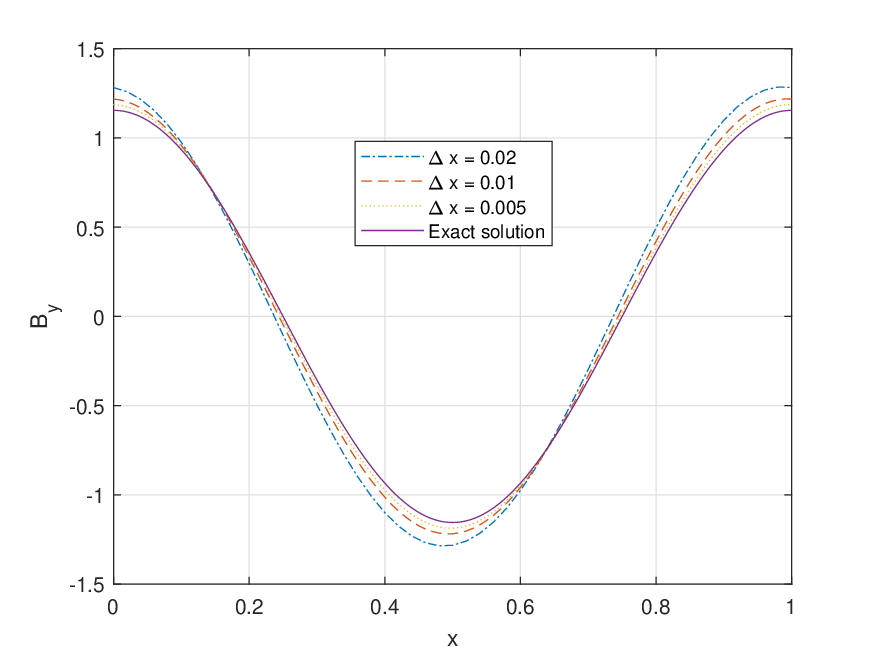}}
  \subfloat[$E^z$]{
  \label{fig:Ey}
  \includegraphics[width=0.49\textwidth]{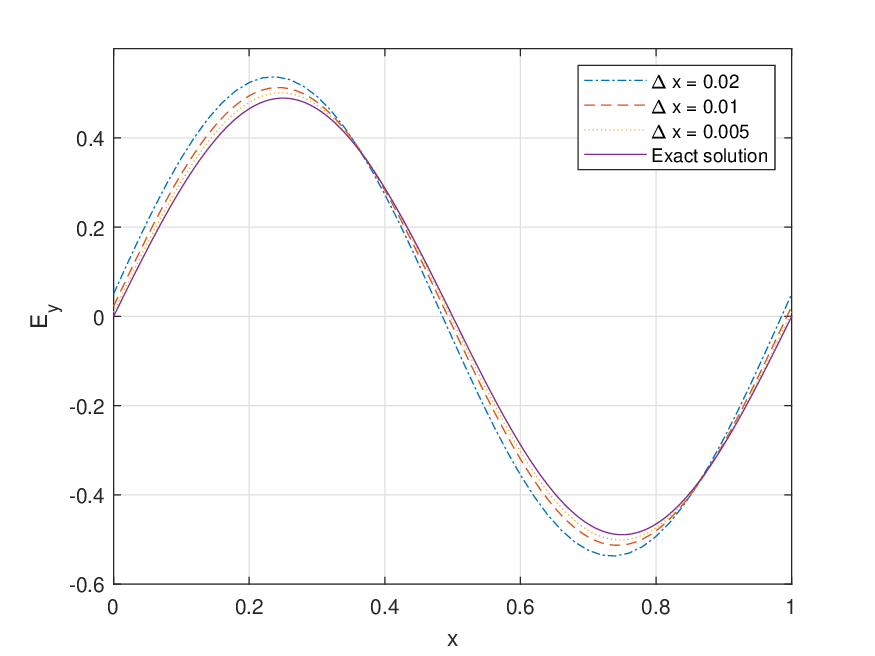}}
 \caption{CP Alfv\'en waves test. Numerical values for $B^y$ and $E^y$ at $t=T$, for three different spatial resolutions, when first-order MIRK method is used. $\sigma=10^8$ and CFL=0.3. The exact solution in the ideal MHD limit is also included.}
\end{figure}

The time coordinate starts at $t=0$ and ends after one period, at $t=T=1/v_A$. In Figures \ref{fig:By} and \ref{fig:Ey} we display the numerical solution of $B^y$ and $E^y$ at $t=T$ for three different resolutions, $\Delta x = 0.02/2^k$, $k=0,1,2$, when a first-order MIRK method is used, together with the exact solution in the ideal MHD limit.

\begin{figure}[t]
\centering
	\subfloat[$B^y$]{
  \label{fig:By2}
  \includegraphics[width=0.49\textwidth]{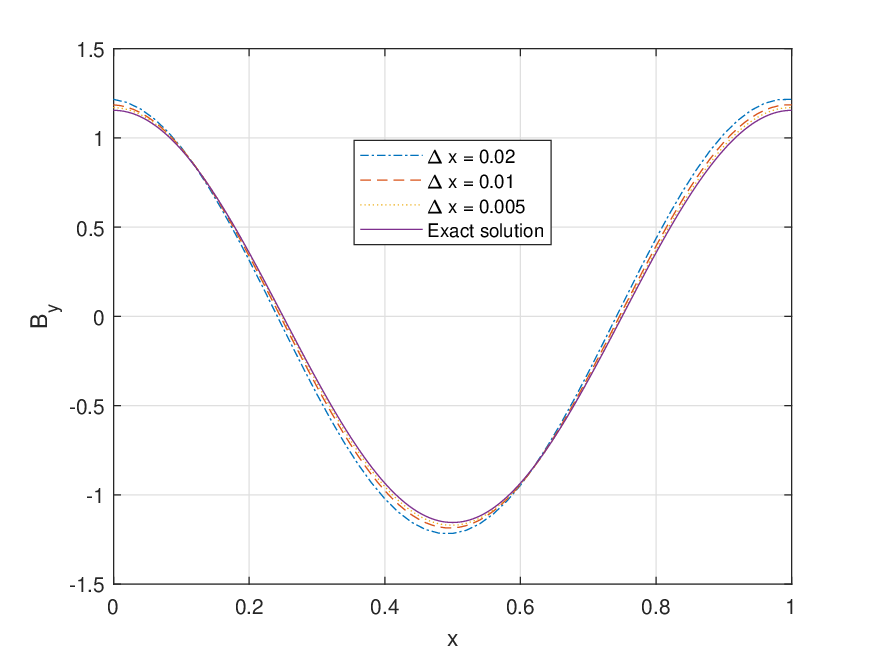}}
  \subfloat[$E^z$]{
  \label{fig:Ey2}
  \includegraphics[width=0.49\textwidth]{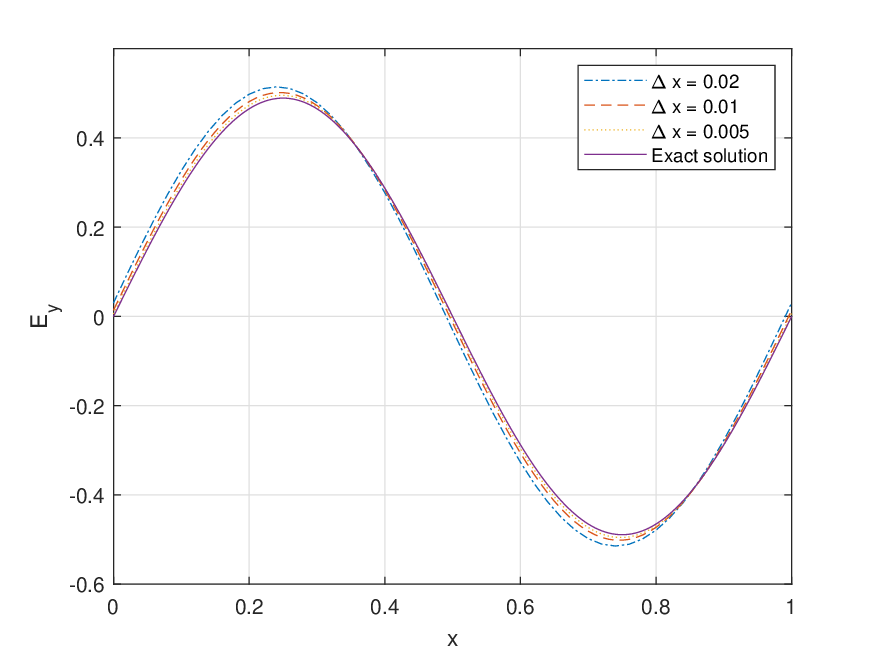}}
 \caption{CP Alfv\'en waves test. Numerical values for $B^y$ and $E^y$ at $t=T$, for three different spatial resolutions, when second-order MIRK method is used. $\sigma=10^8$ and CFL=0.3. The exact solution in the MHD ideal limit is also included.}
\end{figure}

\begin{figure}[t]
\centering
	\subfloat[$B^y$. First-order method.]{
  \label{fig41}
  \includegraphics[width=0.49\textwidth]{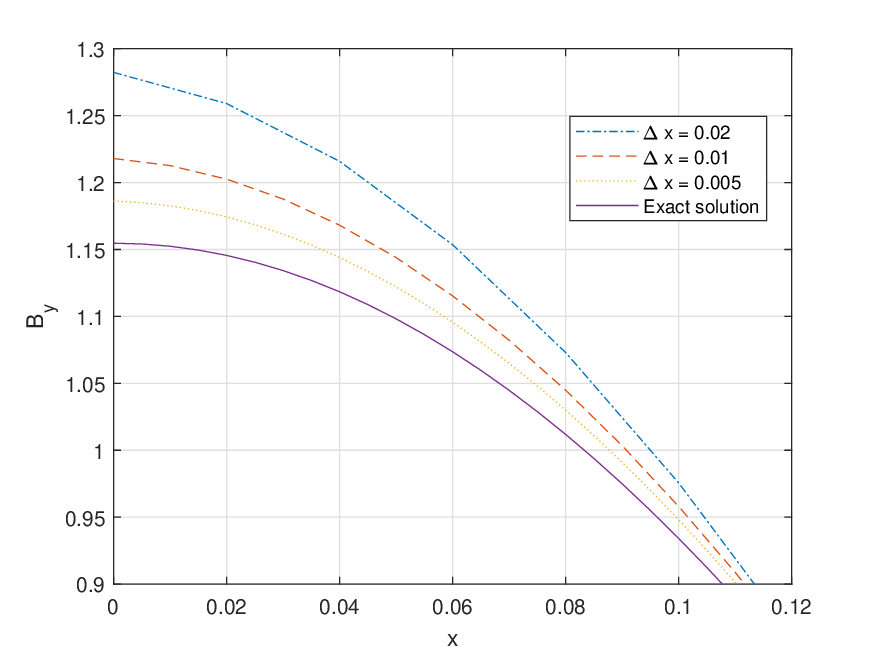}}
  \subfloat[$B^y$. Second-order method.]{
  \label{fig42}
  \includegraphics[width=0.49\textwidth]{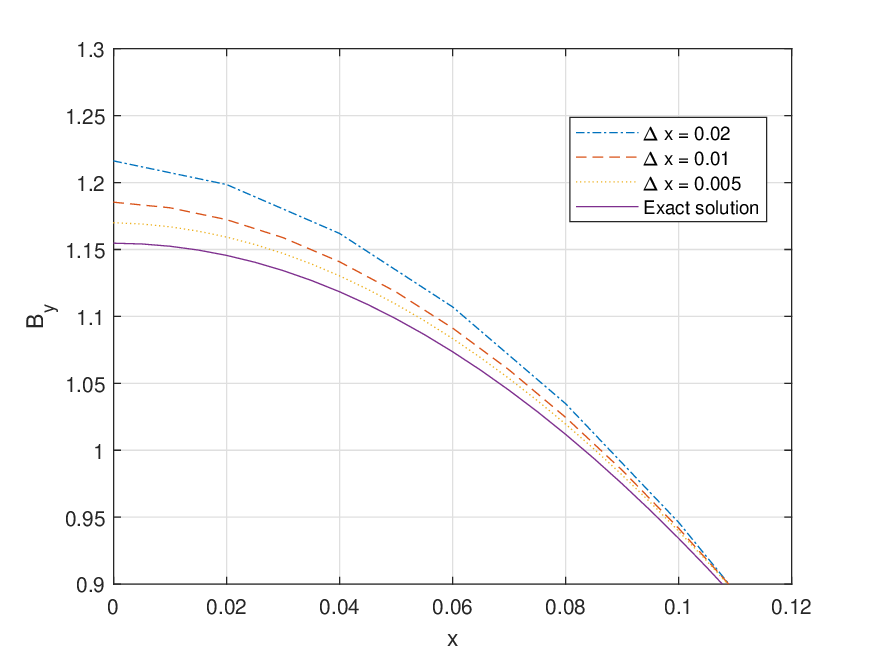}}\\
  \subfloat[$E^y$. First-order method.]{
  \label{fig41}
  \includegraphics[width=0.49\textwidth]{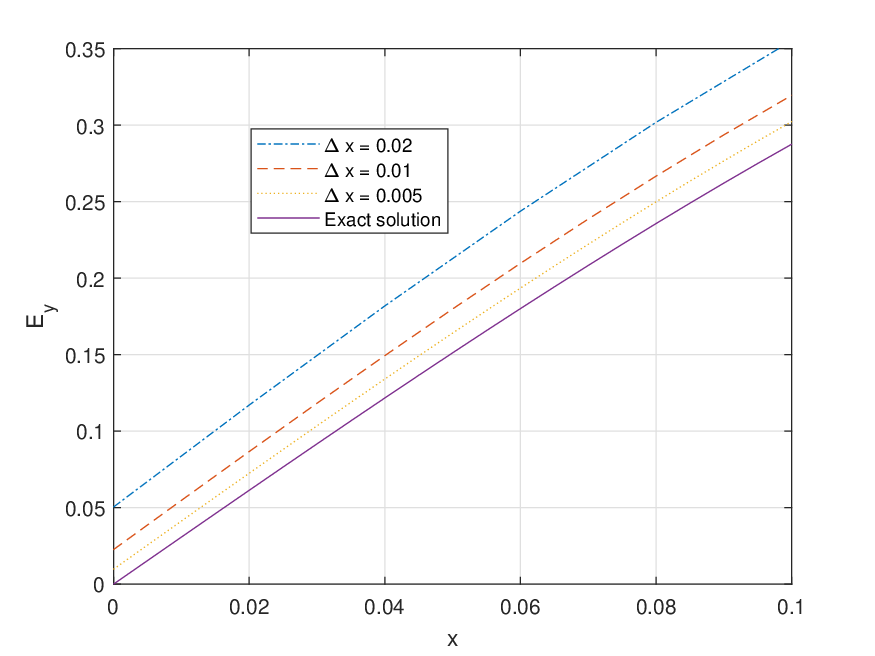}}
  \subfloat[$E^y$. Second-order method.]{
  \label{fig42}
  \includegraphics[width=0.49\textwidth]{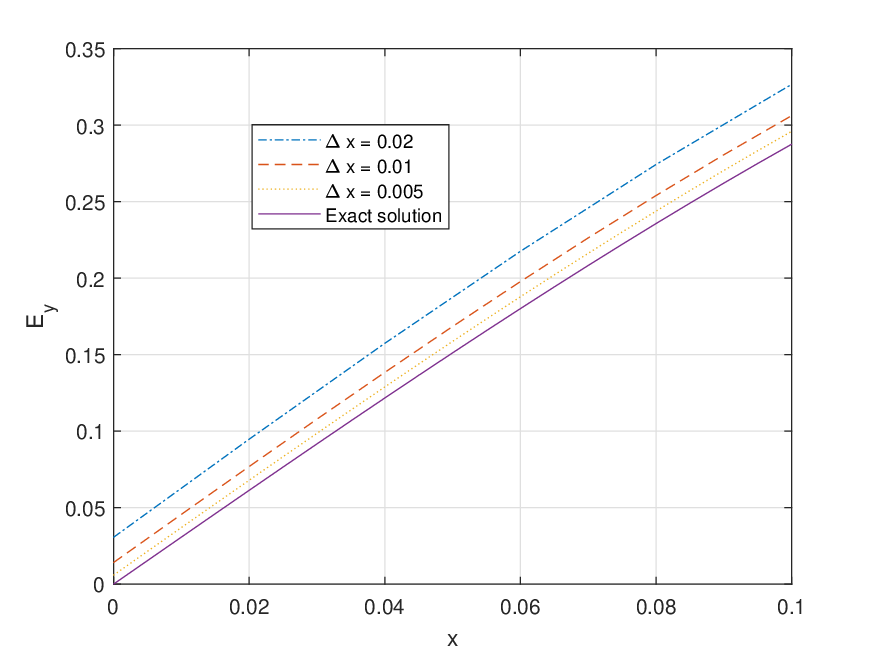}}
 \caption{CP Alfv\'en waves test. Zoom on the numerical values for $B^y$ and $E^y$ at $t=T$, for three different spatial resolutions, when first and second-order MIRK method are used. $\sigma=10^8$ and CFL=0.3. The exact solution in the MHD ideal limit is also included.}
\label{fig:zoom2}
\end{figure}

In the case of second-order MIRK method, we use the same set up as before, but removing the Kreiss-Oliger term (similar results are obtained when this term is not removed). The numerical results are shown in Figures \ref{fig:By2} and \ref{fig:Ey2} for the same variables as for the first-order method. The numerical solution is closer to the exact solution in the ideal MHD limit in comparison with the one obtained with the first-order method, as it can be seen clearer in the zoom of Figures \ref{fig:zoom2}.


We estimate the convergence order of our methods for this test. We carry out simulations for successive smaller resolutions. We have detected that the Kreiss-Oliger dissipation term (\ref{eq:KO}) affects this computation; therefore, we increase in one unit the power of the factor $\Delta x$ and the coefficient $\epsilon=0.1$ in this term. Moreover, we employ a CFL value of 0.1. We consider, as in the previous test, the $L_2$ norm of the errors; in this case, however, since we do not have the exact solution (the exact solution in the ideal MHD limit should be considered as a reference), the error is computed based on the difference of numerical solutions $S$ for successive smaller resolutions, always using points from the coarsest grid:
\begin{equation}
    \varepsilon(\Delta x)=||S(\Delta x)-S(\Delta x /2)||_2.
\label{eq:ord2}
\end{equation}
We get first-order of convergence for the first-order MIRK method as expected, but also first-order of convergence is obtained for the second-order MIRK method. Table \ref{tab:orderalf} shows the obtained results. We were not able to find the reason of this reduction in the order of convergence. Nevertheless, for the second-order MIRK method the orders of convergence computed are always bigger than for the first-order MIRK method. 

\begin{table}[htbp!]
\begin{center}
{\small
\begin{tabular}{|c||c|c|c|c|c|c|}
\hline
$\Delta x$ & $0.04$ & $0.02$ & $0.01$ & $0.005$& $0.0025$ & $0.00125$\\ \hline \hline
  $p$ for the 1st-order & & & & & & \\ MIRK method  & 1.38867 & 0.93344 & 0.87209 & 0.92522 & 0.96129 & 0.98042 \\\hline
  $p$ for the 2nd-order & & & & & & \\ MIRK method &  1.63763 & 1.20525 & 0.99682 & 0.96979 &  0.97859 & 0.98787 \\\hline
\end{tabular}}
\end{center}
\caption{Estimated convergence orders for the first and second-order MIRK methods applied to the CP Alfv\'en waves test at $t=T$, with $\sigma = 10^8$ and CFL = $0.1$, and resolutions $\Delta x=0.04/2^k$, $k=0,1,\ldots,6$, according to Eq. (\ref{eq:ord2}) and the formula (\ref{eq:ord}).}
\label{tab:orderalf}
\end{table}

Finally, we explore three different values for the CFL. We obtain interesting results: for CFL=0.3 and 0.7 we have stable numerical results, but for CFL=0.8 this is no longer the case. This effect is more severe if the Kreiss-Oliger dissipation term is neglected. In Figure \ref{fig:CFL2}, we show the numerical results obtained using first and second-order MIRK methods for these CFL values. It has been checked that the behaviour with and without artificial dissipation in the second-order MIRK method is quite similar.

\begin{figure}[h]
\centering
	\subfloat[First-order method]{
  \includegraphics[width=0.49\textwidth]{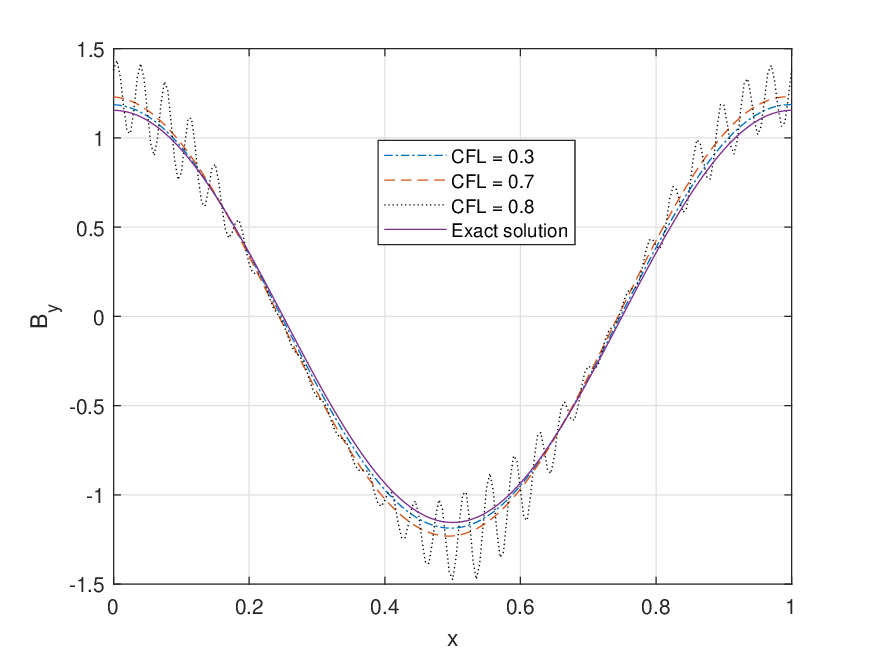}}
  \subfloat[Second-order method without Kreiss-Oliger term]{
  \includegraphics[width=0.49\textwidth]{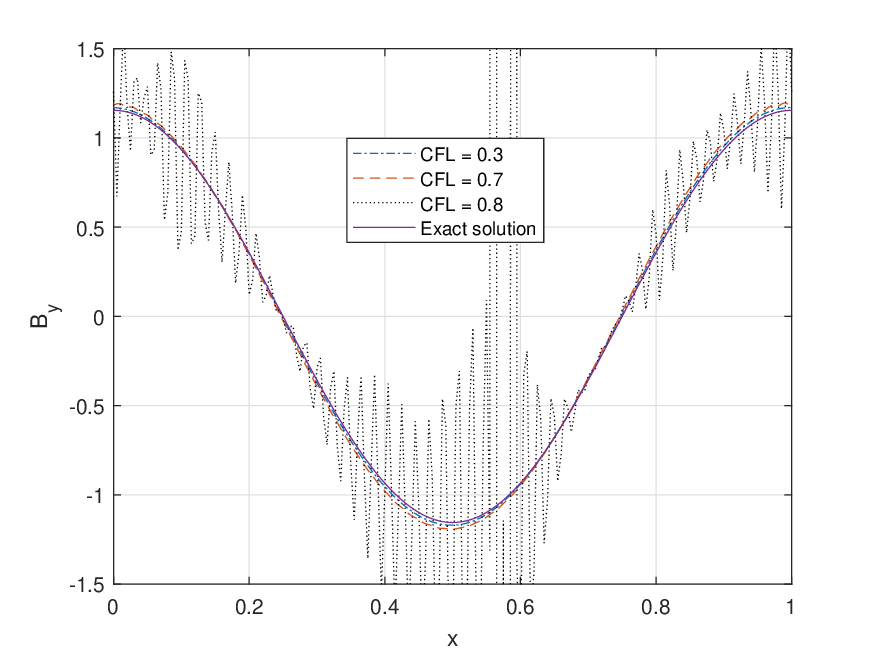}} \\
    \subfloat[Second-order method with Kreiss-Oliger term]{
  \includegraphics[width=0.49\textwidth]{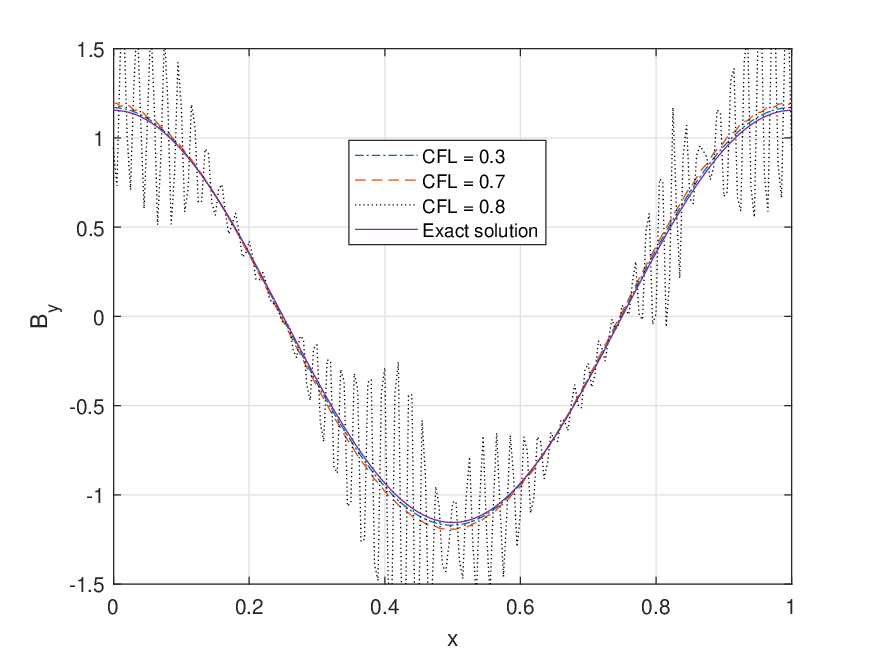}}
 \caption{CP Alfv\'en waves test. Numerical results for several CFL values when first and second-order MIRK methods are used. $\sigma=10^8$ and $\Delta x=0.005$.}
\label{fig:CFL2}
\end{figure}

\section{Conclusions and application in other systems of equations}
In this work, first and second-order MIRK methods have been presented to numerically integrate the RRMHD equations proposed in \cite{Komissarov2007}. In these MIRK methods, only conserved variables are included in the implicit evaluations. The inversion of the operators can be done analytically and is trivial. First and second-order PIRK methods to take into account the wave-like behaviour of the magnetic and electric fields, in addition to linear stability conditions close to the ideal limit, are used to select the values for the $c_i$ coefficients. There is no need of additional iterative steps on each stage with respect to the explicit methods. The potential comparison with IMEX methods strongly depend on the numerical tests carried out: MIRK methods have the same computational cost as explicit ones, while IMEX methods will be more computationally expensive and this cost will depend on the complexity of the equation of state considered (and therefore on the complexity of the recovery process). For both first and second-order MIRK methods, an effective time-step can be defined, making the change of the numerical codes with explicit methods quite direct.

We have also shown some simple dynamical numerical simulations with smooth data, namely the self similar current sheet test and the CP Alfv\'en waves test. More complex simulations and also with non-smooth data are needed to really check the potential of the proposed schemes. Also, the comparison with other approaches, like the ones used in \cite{Palenzuela2009,DumbserZanotti2009}, would be also addressed in future steps. 

The idea behind the MIRK methods can be applied to other kinds of equations. In particular, we are already working on applying this strategy to the numerical resolution of the Boltzmann equations to solve the neutrino transport equations in supernovae simulations using the so-called M1 closure approximation (see reviews mentioned in the introduction for more details). Also, the application to the force-free electrodynamics in its different formulations (see references in \cite{Mahlmann2021}) can be further discussed.

A more general idea is to consider hyperbolic systems of equations with a stiff source term where a parameter can be potentially very large (like the conductivity in the case of RRMHD equations) of the form
\begin{equation}
	\partial_t U + \partial_i F^i(U) = S(U),
\end{equation}
where $U$ is the vector of conserved variables, $F^i$ are the fluxes and the source term $S$ can be written as
\begin{equation}
	S(U) = S_E(U) + \frac{1}{\epsilon} [S_I(U)-U_0];
\end{equation}
here $U_0$ are the values of the conserved variables in the limit $\epsilon\to 0$, $\frac{1}{\epsilon} U_0$ and $S_E$ are explicitly evolved source terms, while $S_I$ is evolved by using MIRK-like methods and can be written as 
\begin{equation}
	S_I(U) = \sum_{i=1}^n G_i(U) \, U^i.
\end{equation}
In the previous expression $G_i$ can depend on the conserved variables $U$ and are always evaluated explicitly; only the components of the vector of conserved variables, $U^i$, multiplying the terms $G_i$, are implicitly evaluated. This means that the inversion of the operators can be done analytically in a very simple way. Moreover, the $c_i$ coefficients appearing when allowing evaluations in different stages of a single time-step should be derived from stability conditions, and the stiff limit can provide key information to select the values for these coefficients.

Identifying if the hyperbolic system of equations for a particular case admits such a decomposition is an art; when this is achieved, one has performed a sort of linearization of the system, quite standard in some physical scenarios. Then, one can study if the application of the MIRK methods provides stable numerical evolutions.

\section*{Acknowledgements}
The numerical work shown in this section has been partially done by C. M.-V. 
during her Final Degree Project, and further developed by the rest of the 
authors of this manuscript. The authors acknowledge support by the Spanish Agencia Estatal de Investigaci\'on / 
Ministerio de Ciencia, Innovaci\'on y Universidades through the Grants No. 
PGC2018-095984-B-I00 and PID2021-125458NB-C21, by the Generalitat Valenciana through the Grants No. 
PROMETEO/2019/071 and ACIF/2019/169 -- European Social Fund. This 
research was partially supported by the Perimeter Institute for Theoretical Physics through the Simons Emmy Noether program. Research at Perimeter Institute is supported by the Government of Canada through the Department of Innovation, Science and Economic Development and by the Province of Ontario through the Ministry of Research and Innovation. The authors acknowledge Luis Lehner and Carlos Palenzuela for fruitful discussions.
						

\begin{thebibliography}{10}
\expandafter\ifx\csname url\endcsname\relax
  \def\url#1{\texttt{#1}}\fi
\expandafter\ifx\csname urlprefix\endcsname\relax\def\urlprefix{URL }\fi
\expandafter\ifx\csname href\endcsname\relax
  \def\href#1#2{#2} \def\path#1{#1}\fi

\bibitem{BalbusHawley}
S.~A. Balbus, J.~F. Hawley, Instability, turbulence, and enhanced transport in
  accretion disks, Reviews of Modern Physics 70 (1998) 1--53.
\newblock \href {https://doi.org/10.1103/RevModPhys.70.1}
  {\path{doi:10.1103/RevModPhys.70.1}}.

\bibitem{BlaesBalbus}
O.~M. Blaes, A.~S. Balbus, Local shear instabilities in weakly ionized, weakly
  magnetized disks, The Astrophysical Journal 421 (1994) 163.
\newblock \href {https://doi.org/10.1086/173634} {\path{doi:10.1086/173634}}.

\bibitem{DeVilliersetal}
J.-P.~D. Villiers, J.~F. Hawley, H.~H. Krolik, Magnetically driven accretion
  flows in the kerr metric. i. models and overall structure, The Astrophysical
  Journal 599 (2003) 1238--1253.
\newblock \href {https://doi.org/10.1086/379509} {\path{doi:10.1086/379509}}.

\bibitem{Fragileetal}
P.~C. Fragile, O.~M. Blaes, P.~Anninos, J.~D. Salmonson, Global general
  relativistic magnetohydrodynamic simulation of a tilted black hole accretion
  disk, The Astrophysical Journal 668 (2007) 417--429.
\newblock \href {https://doi.org/10.1086/521092} {\path{doi:10.1086/521092}}.

\bibitem{Gableretal}
M.~Gabler, P.~Cerd\'a-Dur\'an, J.~A. Font, E.~Mueller, N.~Stergioulas,
  Magneto-elastic oscillations and the damping of crustal shear modes in
  magnetars, Monthly Notices of the Royal Astronomical Society 410 (2011)
  L37--L41.
\newblock \href {https://doi.org/10.1111/j.1745-3933.2010.00974.x}
  {\path{doi:10.1111/j.1745-3933.2010.00974.x}}.

\bibitem{KaspiBeloborodov}
V.~M. Kaspi, A.~M. Beloborodov, Magnetars, Annual Review of Astronomy and
  Astrophysics 55 (2017) 261--301.
\newblock \href {https://doi.org/10.1146/annurev-astro-081915-023329}
  {\path{doi:10.1146/annurev-astro-081915-023329}}.

\bibitem{KomissarovPorth}
S.~Komissarov, O.~Porth, Numerical simulations of jets, New Astronomy Reviews
  92 (2021) 101610.
\newblock \href {https://doi.org/10.1016/j.newar.2021.101610}
  {\path{doi:10.1016/j.newar.2021.101610}}.

\bibitem{Marti}
J.~M. Mart\'i, Numerical simulations of jets from active galactic nuclei,
  Galaxies 7 (2019) 24.
\newblock \href {https://doi.org/10.3390/galaxies7010024}
  {\path{doi:10.3390/galaxies7010024}}.

\bibitem{Perucho}
M.~Perucho, Dissipative processes and their role in the evolution of radio
  galaxies, Galaxies 7 (2019) 70.
\newblock \href {https://doi.org/10.3390/galaxies7030070}
  {\path{doi:10.3390/galaxies7030070}}.

\bibitem{Spruit}
H.~C. Spruit, Dynamo action by differential rotation in a stably stratified
  stellar interior, Astronomy and Astrophysics 381 (2002) 923--932.
\newblock \href {https://doi.org/10.1051/0004-6361:20011465}
  {\path{doi:10.1051/0004-6361:20011465}}.

\bibitem{Palenzuela2009}
C.~Palenzuela, L.~Lehner, O.~Reula, L.~Rezzolla, Beyond ideal {MHD}: towards a
  more realistic modelling of relativistic astrophysical plasmas, Monthly
  Notices of the Royal Astronomical Society 394 (2009) 1727--1740.
\newblock \href {https://doi.org/10.1111/j.1365-2966.2009.14454.x}
  {\path{doi:10.1111/j.1365-2966.2009.14454.x}}.

\bibitem{Komissarov2007}
S.~S. Komissarov, Multidimensional numerical scheme for resistive relativistic
  magnetohydrodynamics, Monthly Notices of the Royal Astronomical Society 382
  (2007) 995--1004.
\newblock \href {https://doi.org/10.1111/j.1365-2966.2007.12448.x}
  {\path{doi:10.1111/j.1365-2966.2007.12448.x}}.

\bibitem{Anton2006}
L.~Ant\'on, O.~Zanotti, J.~A. Miralles, J.~M. Mart\'i, J.~M. Ib\'a{\~n}ez,
  J.~A. Font, J.~A. Pons, Numerical 3+1 general relativistic
  magnetohydrodynamics: a local characteristic approach, The Astrophysical
  Journal 637 (2006) 296.

\bibitem{Mahlmann2021}
J.~F. Mahlmann, M.~A. Aloy, V.~Mewes, P.~Cerd\'a-Dur\'an, Computational general
  relativistic force-free electrodynamics, Astronomy and Astrophysics 647
  (2021) A57.
\newblock \href {https://doi.org/10.1051/0004-6361/202038907}
  {\path{doi:10.1051/0004-6361/202038907}}.

\bibitem{DumbserZanotti2009}
M.~Dumbser, O.~Zanotti, Very high order $p_np_m$ schemes on unstructured meshes
  for the resistive relativistic mhd equations, Journal of Computational
  Physics 228 (2009) 6991--7006.
\newblock \href {https://doi.org/10.1016/j.jcp.2009.06.009}
  {\path{doi:10.1016/j.jcp.2009.06.009}}.

\bibitem{PIRKarxiv}
I.~Cordero-Carri\'on, P.~Cerd\'a-Dur\'an, Partially implicit runge-kutta
  methods for wave-like equations (arxiv:1211.5930) (2012).

\bibitem{PIRKproc}
I.~Cordero-Carri\'on, P.~Cerd\'a-Dur\'an, Partially implicit runge-kutta
  methods for wave-like equations, Advances in Differential Equations and
  Applications, SEMA SIMAI Springer Series (Springer International Publishing
  Switzerland, Switzerland) 4 (2014).

\bibitem{privatePalenzuela}
Private discussion with c. palenzuela.

\end{thebibliography}

\end{document}